\documentclass[a4paper,11pt]{article}
\pdfoutput=1
\usepackage{jheppub}

\usepackage[utf8]{inputenc}

\usepackage{amssymb}
\usepackage{amsmath}
\usepackage{braket}
\usepackage{xcolor}
\usepackage{graphicx}
\usepackage{slashed}

\allowdisplaybreaks

\DeclareMathOperator{\Tr}{Tr}
\DeclareMathOperator{\Porder}{\mathcal{P}}

\newcommand{\e}{\epsilon}

\newcommand{\zero}{{(0)}}
\newcommand{\one}{{(1)}}
\newcommand{\two}{{(2)}}

\newcommand{\als}{\alpha_s}
\newcommand{\alsmu}{\alpha_s(\mu)}
\newcommand{\alspi}{\frac{\alpha_s}{4\pi}}
\newcommand{\alsmupi}{\frac{\alpha_s(\mu)}{4\pi}}

\newcommand{\Ord}{\mathcal{O}}

\newcommand{\nn}{\nonumber}

\newcommand{\df}{d}

\newcommand{\Gcusp}{\Gamma^{\text{cusp}}}
\newcommand{\Lp}{L_\perp}

\newcommand{\kp}{\vec{k}_T}
\newcommand{\kpsq}{k_T^2}
\newcommand{\bp}{\vec{b}_T}
\newcommand{\bpsq}{b_T^2}

\def\cB{\mathcal{B}}
\def\cC{\mathcal{C}} 
\def\cD{\mathcal{D}}

\def\cI{\mathcal{I}}

\def\cO{\mathcal{O}}
\def\cS{\mathcal{S}}

\preprint{}

\title{Transverse Parton Distribution and Fragmentation Functions at NNLO: the Quark Case}

\author[a]{Ming-Xing Luo,}
\emailAdd{mingxingluo@zju.edu.cn}
\author[b]{Xing Wang,}
\emailAdd{x.wong@pku.edu.cn}
\author[b]{Xiaofeng Xu,}
\emailAdd{xuxiaofeng@pku.edu.cn}
\author[b,c]{Li Lin Yang,}
\emailAdd{yanglilin@pku.edu.cn}
\author[a]{Tong-Zhi Yang,}
\emailAdd{yangtz@zju.edu.cn}
\author[a]{and Hua Xing Zhu}
\emailAdd{zhuhx@zju.edu.cn}

\affiliation[a]{Zhejiang Institute of Modern Physics, Department of Physics, Zhejiang University, Hangzhou, 310027, China}
\affiliation[b]{School of Physics and State Key Laboratory of Nuclear Physics and Technology, Peking University, Beijing 100871, China}
\affiliation[c]{Center for High Energy Physics, Peking University, Beijing 100871, China}

\abstract{We revisit the calculation of perturbative quark transverse momentum dependent parton distribution functions and fragmentation functions using the exponential regulator for rapidity divergences. We show that the exponential regulator provides a consistent framework for the calculation of various ingredients in transverse momentum dependent factorization. Compared to existing regulators in the literature, the exponential regulator has a couple of advantages which we explain in detail. As a result, the calculation is greatly simplified and we are able to obtain the next-to-next-to-leading order results up to $\Ord(\e^2)$ in dimensional regularization. These terms are necessary for a higher order calculation which is made possible with the simplification brought by the new regulator. As a by-product, we have obtained the two-loop quark jet function for the Energy-Energy Correlator in the back-to-back limit, which is the last missing ingredient for its N$^3$LL resummation.}


\begin{document}

\maketitle

\clearpage

\section{Introduction}
\label{sec:introduction}

Parton distribution functions (PDFs) and fragmentation functions (FFs) describe the partonic contents of hadrons. They are of fundamental importance in quantum chromodynamics (QCD)~\cite{Collins:2011zzd,Angeles-Martinez:2015sea}. They enter factorization formulas for scattering processes involving hadrons, and are essential for comparing theoretical predictions for the cross sections against experimental measurements. In most cases, the observables are only sensitive to the longitudinal momenta of the partons, and the transverse momenta can be integrated over, leading to the so-called ``collinear'' PDFs and FFs. However, in certain regions of phase space, the transverse momenta of the partons become relevant, and one needs the transverse momentum dependent PDFs (TMDPDFs) and FFs (TMDFFs) in the corresponding factorization formulas. This is case for the small transverse momentum ($Q_T$) region in the Drell-Yan process~\cite{Dokshitzer:1978yd,Parisi:1979se,Collins:1984kg,Arnold:1990yk,Ladinsky:1993zn,Bozzi:2010xn,Becher:2011xn,Bizon:2019zgf,Bertone:2019nxa}, and also for similar regions in, e.g., semi-inclusive deep-inelastic scattering (SIDIS)~\cite{Ji:2004wu,Ji:2004xq,Su:2014wpa,Kang:2015msa,Liu:2018trl}, electron-positron annihilation to hadrons and jets~\cite{Collins:1981uk,Collins:1981va,Neill:2016vbi,Gutierrez-Reyes:2018qez,Gutierrez-Reyes:2019vbx,Gutierrez-Reyes:2019msa}, Higgs boson production~\cite{Berger:2002ut,Bozzi:2005wk,Gao:2005iu,Echevarria:2015uaa,Neill:2015roa,Bizon:2017rah,Chen:2018pzu,Bizon:2018foh}, top quark pair production~\cite{Zhu:2012ts,Li:2013mia,Catani:2014qha,Catani:2018mei}, as well as Energy-Energy Correlator (EEC) in the back-to-back limit at both lepton and hadron colliders~\cite{Moult:2018jzp,Gao:2019ojf}. To improve the theoretical predictions for these observables, it is desirable to have precise knowledges about these basic objects. 

TMDPDFs and TMDFFs can be defined as hadronic matrix elements of bilinear quark or gluon field operators with a measured transverse momentum $\vec{k}_\perp$ (or a transverse separation $\vec{b}_\perp$ in position space). If the transverse momentum $\vec{k}_\perp \sim \Lambda_{\text{QCD}}$, the TMDPDFs and TMDFFs are essentially non-perturbative, and can only be extracted from experimental data or calculated using lattice methods. On the other hand, if $\vec{k}_\perp \gg \Lambda_{\text{QCD}}$, the TMDPDFs and TMDFFs can be related to the collinear PDFs and FFs via perturbatively calculable matching coefficients. These coefficients are known at the next-to-next-to-leading order (NNLO) for the TMDPDFs~\cite{Catani:2011kr,Gehrmann:2012ze,Gehrmann:2014yya,Echevarria:2016scs} and TMDFFs~\cite{Echevarria:2016scs}. They have played an important role in a number of cutting-edge calculations, including precision predictions for the Drell-Yan process and Higgs boson production at small transverse momentum \cite{Chen:2018pzu,Bizon:2018foh,Bizon:2019zgf}, and NNLO calculations for top quark pair production using the $Q_T$ subtraction method~\cite{Catani:2019iny,Catani:2019hip}.

In this work, we revisit the calculation of the matching coefficients for TMDPDFs and TMDFFs at NNLO. We consider the quark TMDPDFs and TMDFFs in this paper, while the gluon case is left to a forthcoming article. There are several new elements in our calculation compared to those in the literature:
\begin{itemize}

\item We employ the exponential regulator for rapidity divergences~\cite{Li:2016axz}. Rapidity divergences or ``collinear anomalies'' appear in the calculation of individual TMD functions in factorization formulas, which are cancelled in physical observables. These divergences are not regularized by dimensional regularization, and additional regulators need to be introduced~\cite{Collins:1984kg,Ji:2004wu,Collins:2011zzd,Becher:2010tm,Becher:2011dz,Chiu:2012ir,Chiu:2009yx,Echevarria:2015byo,Ebert:2018gsn}. 
The exponential regulator has been shown to be particularly suitable in the calculation of TMD soft functions, as demonstrated in the recent next-to-next-to-next-to-leading order (N$^3$LO) calculation~\cite{Li:2016ctv}. We show in this work that the exponential regulator can also be used to calculate TMDPDFs and TMDFFs, which are more complicated objects than TMD soft functions.
Our results show that the exponential regulator is a consistent rapidity regulator in both the soft and collinear sectors.

\item We develop systematic calculation method based on modern techniques for loop integrals, such as integration-by-parts~(IBP) identities~\cite{Chetyrkin:1981qh,Laporta:2001dd} and differential equations~\cite{Bern:1993kr,Gehrmann:1999as,Henn:2013pwa}. Our method paves the way to calculate TMDPDFs and TMDFFs at N$^3$LO. 

\item We obtain the bare NNLO TMDPDFs and TMDFFs up to $\mathcal{O}(\epsilon^2)$, where $\epsilon$ is the dimensional regulator. They directly contribute to TMDPDFs and TMDFFs at N$^3$LO upon renormalization.

\item Our results for TMDPDFs agree with previous calculations~\cite{Gehrmann:2012ze,Gehrmann:2014yya,Echevarria:2016scs}, but we find a small discrepancy for the TMDFFs compared to those presented in Ref.~\cite{Echevarria:2016scs}. We have performed several consistency checks on our results to make sure that they are correct.

\item As a by-product, we obtain the NNLO quark jet function relevant for the resummation of EEC in the back-to-back limit. This is the last missing ingredient for this resummation at the next-to-next-to-next-to-leading logarithmic (N$^3$LL) accuracy.

\end{itemize}

This paper is organized as follows. In Section~\ref{sec:fac} we introduce the definitions of quark TMDPDFs and TMDFFs in the context of the SIDIS process, and discuss the exponential regulator for rapidity divergences. In Section~\ref{sec:tmdpdf} and \ref{sec:tmdff} we perform the calculation of the quark TMDPDFs and TMDFFs at NNLO using the exponential regulator. In Section~\ref{sec:tmdff} we also use the results for TMDFFs to compute the two-loop jet function for EEC in the back-to-back limit. This is by itself a new result of our paper, and also serves as a cross-check of our results. We conclude in Section~\ref{sec:conclusion}.

\section{Transverse momentum dependent factorization}
\label{sec:fac}

\subsection{Kinematics and factorization}

In this section, we briefly review the formalism of transverse momentum dependent factorization and introduce the definitions of TMDPDFs and TMDFFs. For our purpose, it is easiest to consider (unpolarized) SIDIS which involves hadrons in both the initial state and the final state. In SIDIS, a hadron $N_1$ with momentum $P_1^\mu$ is probed by a virtual photon $\gamma^*$ with momentum $q^\mu$ and produces a jet containing a specific hadron $N_2$ with momentum $P_2^\mu$. We define the kinematic invariants
\begin{equation}
Q^2 \equiv -q^2 \, , \quad x \equiv \frac{Q^2}{2P_1 \cdot q} \, , \quad z \equiv \frac{P_1 \cdot P_2}{P_1 \cdot q} \, .
\end{equation}
We introduce two light-like 4-vectors $n$ and $\bar{n}$ satisfying $n^2 = \bar{n}^2 = 0$ and $n \cdot \bar{n} = 2$, such that we can decompose any 4-vector $k^\mu$ as
\begin{equation}
k^\mu = \bar{n} \cdot k \, \frac{n^\mu}{2} + n \cdot k \, \frac{\bar{n}^\mu}{2} + k_\perp^\mu \equiv k_+ \, \frac{n^\mu}{2} + k_- \, \frac{\bar{n}^\mu}{2} + k_\perp^\mu \, .
\end{equation}
When quoting the components of a 4-vector, we use $k = (k_+, k_-, k_\perp)$. The scalar product of two 4-vectors is given by
\begin{equation}
p \cdot k = \frac{p_+ k_- + p_- k_+}{2} + p_\perp \cdot k_\perp \, .
\end{equation}
In the hadron frame and ignoring the hadron masses, we have
\begin{equation}
P_1 = (P_{1+}, 0 , 0_\perp) \, , \quad P_2 = (0, P_{2-}, 0_\perp) \, , \quad q = (q_+,q_-,q_\perp) \, ,
\end{equation}
where
\begin{equation}
q_- = \frac{P_{2-}}{z} = \frac{Q^2}{x P_{1+}} \, , \quad q_+q_- + q_\perp^2 = -Q^2 \, ,
\end{equation}
and we define $q_\perp^2 \equiv -q_T^2$.

The hadronic tensor is defined as
\begin{multline}
W^{\mu\nu} \equiv \sum_X (2\pi)^4 \delta^{(4)}(P_1+q-P_2-P_X)
\\
\times \braket{N_1(P_1) | J^\mu(0) | N_2(P_2),X} \braket{N_2(P_2),X | J^\nu(0) | N_1(P_1)} \, .
\end{multline}
In the region $q_T \sim Q \gg \Lambda_{\text{QCD}}$, the hadronic tensor can be factorized into products of hard kernels with collinear PDFs and FFs:
\begin{equation}
W^{\mu\nu} = \sum_{i,j} H^{\mu\nu}_{ij}(Q,q_\perp,\mu) \, \phi_{i/N_1}(x,\mu) \, d_{N_2/j}(z,\mu) + \mathcal{O}(\Lambda_{\text{QCD}}^2/Q^2) \, ,
\end{equation}
where we have suppressed the dependence of the hard kernel on other kinematic variables.
In the language of soft-collinear effective theory (SCET) \cite{Bauer:2000ew,Bauer:2000yr,Bauer:2001yt,Bauer:2002nz,Beneke:2002ph}, the collinear PDFs and FFs can be defined as matrix elements of gauge-invariant collinear fields. For example, the bare quark collinear PDF and FF are defined by \cite{Collins:1981uw,Bauer:2002nz,Collins:2011zzd}
\begin{align}
\phi^{\text{bare}}_{q/N_1}(x) &= \int \frac{dt}{2\pi} \, e^{-i x t \bar{n} \cdot P_1} \, \braket{N_1(P_1) | \bar{\chi}_n(t\bar{n}) \frac{\slashed{\bar{n}}}{2} \chi_n(0) | N_1(P_1)} \, , \nonumber
\\
d^{\text{bare}}_{N_2/q}(z) &= \sum_X z^{1-2\epsilon} \int \frac{dt}{2\pi} \, e^{i t n \cdot P_2 / z} \, \Tr \braket{0 | \frac{\slashed{n}}{2} \chi_{\bar{n}}(tn) | N_2(P_2),X} \braket{N_2(P_2),X | \bar{\chi}_{\bar{n}}(0) | 0} \, ,
\label{eq:collinear_pdf_ff}
\end{align}
where $\chi_n$ and $\chi_{\bar{n}}$ are the gauge-invariant collinear quark fields along the $n$ and $\bar{n}$ directions, respectively. We have assumed dimensional regularization with $d=4-2\epsilon$. The collinear PDF $\phi_{q/N_1}(x,\mu)$ describes (in a sense) the probability distribution of finding the quark $q$ with momentum fraction $x$ inside the fast-moving hadron $N_1$. The collinear FF $d_{N_2/q}(z,\mu)$, on the other hand, describes the probability distribution of finding the hadron $N_2$ with momentum fraction $z$ inside the jet initiated by the quark $q$.

If $q_T \ll Q$, however, the above picture of collinear factorization breaks down due to the appearance of large logarithms of $q_T/Q$ in the hard kernel $H_{ij}$. One should instead rely on TMD factorization of the form
\begin{align}
W^{\mu\nu} &= \sum_{i,j} H^{\prime\mu\nu}_{ij}(Q,\mu) \int \frac{d^2b_\perp}{(2\pi)^2} \, e^{ib_\perp \cdot q_\perp} \nonumber
\\
&\hspace{4em} \times \mathcal{B}_{i/N_1}(x,b_\perp,\mu) \, \mathcal{D}_{N_2/j}(z,b_\perp,\mu) \, \mathcal{S}_{ij}(b_\perp,\mu) + \mathcal{O}(q_T^2/Q^2) \nonumber
\\
&= \sum_{i,j} H^{\prime\mu\nu}_{ij}(Q,\mu) \int d^2k_{1\perp} d^2k_{2\perp} d^2k_{s\perp} \, \delta^{(2)}(k_{1\perp}+q_\perp-k_{2\perp}-k_{s\perp}) \nonumber
\\
&\hspace{4em} \times \tilde{\mathcal{B}}_{i/N_1}(x,k_{1\perp},\mu) \, \tilde{\mathcal{D}}_{N_2/j}(z,k_{2\perp},\mu) \, \tilde{\mathcal{S}}_{ij}(k_{s\perp},\mu) + \mathcal{O}(q_T^2/Q^2) \, ,
\label{eq:TMDfac}
\end{align}
where $\mathcal{B}_{i/N_1}$, $\mathcal{D}_{N_2/j}$ and $\mathcal{S}_{ij}$ are TMDPDFs, TMDFFs and TMD soft functions in the impact parameter space, with $b_\perp$ the impact parameter; while $\tilde{\mathcal{B}}_{i/N_1}$, $\tilde{\mathcal{D}}_{N_2/j}$ and $\tilde{\mathcal{S}}_{ij}$ are their counterparts in the transverse momentum space. For our purpose, we only consider $i,j$ being quarks and anti-quarks.

\begin{figure}[t!]
\centering
\includegraphics[width=0.45\linewidth]{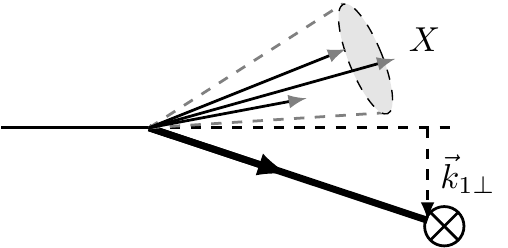}
\hspace{1em}
\includegraphics[width=0.4\linewidth]{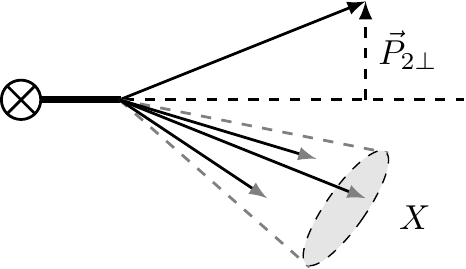}
\caption{Kinematics for TMDPDFs (left plot) and for TMDFFs in the parton frame (right plot).}
\label{fig:TMD_PDF_FF}
\end{figure}

The quark TMDPDF $\tilde{B}_{q/N_1}(x,k_{1\perp},\mu)$ describes the probability distribution of find a quark with momentum fraction $x$ and transverse momentum $k_{1\perp}$ inside the hadron $N_1$, as depicted in the left plot of Figure~\ref{fig:TMD_PDF_FF}. Naively, the bare quark TMDPDF can be defined by
\begin{align}
\mathcal{B}^{\text{bare}}_{q/N_1}(x,b_\perp) &\equiv \int d^{d-2}k_{1\perp} \, e^{i b_\perp \cdot k_{1\perp}} \, \tilde{B}^{\text{bare}}_{q/N_1}(x,k_{1\perp}) \nonumber
\\
&\equiv \int \frac{dt}{2\pi} \, e^{-i x t \bar{n} \cdot P_1} \, \braket{N_1(P_1) | \bar{\chi}_n(t\bar{n}+b_\perp) \frac{\slashed{\bar{n}}}{2} \chi_n(0) | N_1(P_1)} \nonumber
\\
&= \int \frac{db_-}{4\pi} \, e^{-i x b_- P_{1+}/2} \, \braket{N_1(P_1) | \bar{\chi}_n(0,b_-,b_\perp) \frac{\slashed{\bar{n}}}{2} \chi_n(0) | N_1(P_1)} \, .
\label{eq:TMDPDFnaive}
\end{align}
Similarly, the bare quark TMDFF may be defined as
\begin{align}
\mathcal{D}^{\text{bare}}_{N_2/q}(z,b_\perp) &\equiv \int d^{d-2}k_{2\perp} \, e^{-ib_\perp \cdot k_{2\perp}} \, \tilde{\mathcal{D}}^{\text{bare}}_{N_2/q}(z,k_{2\perp}) \nonumber
\\
&\hspace{-3em} \equiv \sum_X \frac{1}{z} \int \frac{dt}{2\pi} \, e^{i t n \cdot P_2 / z} \, \Tr \braket{0 | \frac{\slashed{n}}{2} \chi_{\bar{n}}(tn+b_\perp) | N_2(P_2),X} \braket{N_2(P_2),X | \bar{\chi}_{\bar{n}}(0) | 0} \nonumber
\\
&\hspace{-3em} = \sum_X \frac{1}{z} \int \frac{db_+}{4\pi} \, e^{i b_+ P_{2-} / (2z)} \, \Tr \braket{0 | \frac{\slashed{n}}{2} \chi_{\bar{n}}(b_+,0,b_\perp) | N_2(P_2),X} \braket{N_2(P_2),X | \bar{\chi}_{\bar{n}}(0) | 0} \, .
\label{eq:TMDFFnaive}
\end{align}
Note that in the above definition, $k_{2\perp}$ represents the transverse momentum of the quark in the hadron frame (where $N_2$ has zero transverse momentum). In practice, it is also useful to define the TMDFFs in the parton frame where the quark has zero transverse momentum. In the parton frame, $N_2$ now has a non-zero transverse momentum $P_{2\perp}$ which is related to $k_{2\perp}$ by $P_{2\perp} = -z k_{2\perp}$. The parton frame quark TMDFF is then
\begin{equation}
\mathcal{F}^{\text{bare}}_{N_2/q}(z,b_\perp/z) \equiv \int d^{d-2}P_{2\perp} \, e^{ib_\perp \cdot P_{2\perp} / z} \, \tilde{\mathcal{F}}^{\text{bare}}_{N_2/q}(z,P_{2\perp}) \, ,
\end{equation}
with
\begin{multline}
\tilde{\mathcal{F}}^{\text{bare}}_{N_2/q}(z,P_{2\perp}) = \sum_X \frac{1}{z} \int \frac{db_+}{4\pi} \frac{d^{d-2}b_\perp}{(2\pi)^{d-2}} \, e^{i b_+ P_{2-} / (2z)}
\\
\times \Tr \braket{0 | \frac{\slashed{n}}{2} \chi_{\bar{n}}(b_+,0,b_\perp) | N_2(P_2),X} \braket{N_2(P_2),X | \bar{\chi}_{\bar{n}}(0) | 0} \, .
\end{multline}
Here, $P_{2\perp}$ is defined with respect to the axis chosen such that the total transverse momentum of $N_2$ and $X$ is zero. It is easy to show that
\begin{equation}
\tilde{\mathcal{F}}^{\text{bare}}_{N_2/q}(z,P_{2\perp}) = \tilde{\mathcal{D}}^{\text{bare}}_{N_2/q}(z,-P_{2\perp}/z) \, .
\label{eq:ff1_parton_hadron}
\end{equation}
The function $\tilde{\mathcal{F}}^{\text{bare}}_{N_2/q}(z,P_{2\perp})$ represents the probability distribution of finding a hadron $N_2$ with momentum fraction $z$ and transverse momentum $P_{2\perp}$ inside the jet initiated by the quark $q$, as depicted in the right plot of Figure~\ref{fig:TMD_PDF_FF}. From the above definitions, it is easy to see that
\begin{equation}
\mathcal{F}^{\text{bare}}_{N_2/q}(z,b_\perp/z) = z^{2-2\epsilon} \, \mathcal{D}^{\text{bare}}_{N_2/q}(z,b_\perp) \, .
\label{eq:ff2_parton_hadron}
\end{equation}
Finally, the quark TMD soft function is given by the vacuum expectation value of a soft Wilson loop
\begin{equation}
\mathcal{S}^{\text{bare}}_{q\bar{q}}(b_\perp) \equiv \frac{1}{N_c} \Tr \braket{0 | S^\dagger_{\bar{n}}(b_\perp) \, S_n(b_\perp) \, S_n^\dagger(0) \, S_{\bar{n}}(0) | 0} \, ,
\label{eq:TMDsoftNaive}
\end{equation}
where the soft Wilson line is defined by
\begin{equation}
S_n(x) \equiv \Porder \exp \bigg( ig_s \int_{-\infty}^0 ds \, n \cdot A_s(x+sn) \bigg) \, ,
\end{equation}
with $A_s$ the soft gluon field in SCET.

\subsection{Rapidity divergences and the exponential regulator}

While the TMD factorization formula \eqref{eq:TMDfac} makes some sense, the TMDPDF \eqref{eq:TMDPDFnaive}, TMDFF \eqref{eq:TMDFFnaive} and TMD soft function \eqref{eq:TMDsoftNaive} are actually ill-defined due to the appearance of rapidity divergences which are not regularized in dimensional regularization. These divergences cancel when one combines the 3 functions in the factorization formula \eqref{eq:TMDfac} to calculate physical observables. However, they also carry important information, just like the relationship between ultraviolet (UV) divergences and the renormalization group.

The rapidity divergences arise due to the fact that the collinear modes and soft modes have the same typical off-shellness around $q_T^2$. More precisely, in the $q_T \ll Q$ limit we have the relevant momentum regions
\begin{align}
\label{eq:modes}
\text{collinear: } p_n &\sim Q(1,\lambda^2,\lambda) \, , \nonumber
\\
\text{anti-collinear: } p_{\bar{n}} &\sim Q(\lambda^2,1,\lambda) \, , \nonumber
\\
\text{soft: } p_s & \sim Q(\lambda,\lambda,\lambda) \, .	
\end{align}
where $\lambda=q_T/Q \ll 1$. The effective field theory describing these modes are sometimes called $\text{SCET}_{\text{II}}$. The collinear modes and the soft mode are related by a boost in the $n$ or $\bar{n}$ direction. As a result, they cannot really be separated by a boost invariant regulator such as dimensional regularization. A brute-force separation as done in Eq.~\eqref{eq:TMDfac} then leads to inconsistencies manifesting themselves as rapidity divergences.

To deal with the rapidity divergences, one needs to introduce a regulator in addition to dimensional regularization. This however leads to another subtle issue. Any such regulator necessarily reintroduces a logarithmic dependence on the hard scale $Q$ into integrals in the collinear and anti-collinear regions through $\bar{n} \cdot P_1$ and $n \cdot P_2$, which was supposed to be factorized out into the hard function $H'$ in Eq.~\eqref{eq:TMDfac}. This fact is sometimes called ``collinear anomaly'' or ``factorization anomaly'' in the literature \cite{MartinBeneke, Becher:2010tm}. Nevertheless, using the structure of the rapidity divergences, it can be shown that these $Q$-dependence can be extracted and exponentiated to all orders. After such a ``re-factorization''
\begin{multline}
\mathcal{B}_{q/N_1}(x,b_\perp,\mu) \, \mathcal{D}_{N_2/q}(z,b_\perp,\mu) \, \mathcal{S}_{q\bar{q}}(b_\perp,\mu)
\\
= \left( \frac{b_T^2 Q^2}{b_0^2} \right)^{-F_{q\bar{q}}(b_\perp,\mu)} \, B_{q/N_1}(x,b_\perp,\mu) \, D_{N_2/q}(z,b_\perp,\mu) \, ,
\label{eq:refac}
\end{multline}
where
\begin{equation}
b_T^2=-b_\perp^2 \, , \quad b_0=2e^{-\gamma_E} \, .
\end{equation}
The functions $B_{q/N_1}$ and $D_{N_2/q}$ can be regarded as the ``genuine'' quark TMDPDF and TMDFF which are free from rapidity divergences and are also independent of $Q$. The exponent function $F_{q\bar{q}}$ is closely related to the so-called Collins-Soper kernel~\cite{Collins:2011zzd}. It has been known perturbatively to three loops~\cite{Li:2016ctv,Vladimirov:2016dll}. Very recently, there are proposals to compute it non-perturbatively on the lattice~\cite{Ebert:2018gzl,Ebert:2019okf}.

In the literature, there are a variety of ways to regularize the rapidity divergences \cite{Ji:2004wu,Chiu:2009yx,Collins:2011zzd,Chiu:2012ir,Becher:2011dz,Echevarria:2015byo,Ebert:2018gsn}. In this paper, we consider the so-called exponential regulator \cite{Li:2016axz} which was used to calculate the TMD soft function to the N$^3$LO. We will show that it is a consistent regularization scheme also for the TMDPDFs and TMDFFs. Before discussing the exponential regulator, we briefly review the $\eta$-regulator of Ref.~\cite{Chiu:2012ir} which shares many similarities. At the next-to-leading order (NLO), the $\eta$-regulator amounts to the subsitution
\begin{align}
\label{eq:eta_reg}
\int\frac{\df^dk}{(2\pi)^{d}} \, (2\pi) \delta_+(k^2) \to   \int\frac{\df^dk}{(2\pi)^d} \frac{\nu^{2\eta}}{|2k_z|^{2\eta}} \, (2\pi) \delta_+(k^2) \, ,
\end{align}
for the phase-space integrals over the real gluon momentum $k^\mu$, where $\delta_+(k^2) = \theta(k^0)\delta(k^2)$. The rapidity divergences appear as $1/\eta$ poles which can be subtracted in the same way as renormalizing the UV divergences. After the subtraction, the TMDPDFs, TMDFFs and TMD soft functions still depend on the ``rapidity scale'' $\nu$. For the TMDPDFs and TMDFFs, the natural rapidity scale is $\nu \sim Q$, while for the TMD soft functions $\nu \sim q_T$. The evolution equations of these functions with respect to $\nu$ can be used to exponentiate the rapidity logarithms $\ln(b_T^2Q^2)$ leading to the refactorization in Eq.~\eqref{eq:refac}.

While the above $\eta$-regulator is conceptually simple, it is not easy to implement in higher order calculations beyond NLO. For example, the regulator has to be carefully applied to maintain non-Abelian exponentiation in the soft sector~\cite{Chiu:2012ir,Luebbert:2016itl}. In particular, when there are two real gluon emissions with momenta $k_1$ and $k_2$, it is different to apply the regulator on $k_{1z}+k_{2z}$ as a whole, or on $k_{1z}$ and $k_{2z}$ separately. Recently, a new regulator for rapidity divergences called ``exponential regulator'' has been proposed in Ref.~\cite{Li:2016axz}, which leads to the same rapidity evolution equations as the $\eta$-regulator, and is easier for higher order calculations. In momentum space, the new rapidity regulator is simply multiplying each
soft/collinear phase space measure by an exponential factor
\begin{align}
\int \frac{d^dk}{(2\pi)^d} \, (2\pi) \delta_+(k^2) \to
\lim_{\tau \to 0} \int\frac{d^dk}{(2\pi)^d} \, (2\pi) \delta_+(k^2) \,
\exp(-b_0 \tau k^0) \, .
\end{align}
Note that the $\tau \to 0$ limit has to be taken after integration. Beyond NLO,
when there are multiple soft/collinear partons, the regularization simple
becomes
\begin{align}
\prod_{i=1}^n \int \frac{d^dk_i}{(2\pi)^d} \, (2\pi) \delta_+(k_i^2) \to
\lim_{\tau \to 0} \prod_{i=1}^n \int \frac{d^dk_i}{(2\pi)^d} \, (2\pi) \delta_+(k_i^2) \,
\exp \left( -b_0 \tau \sum_i^n k_i^0 \right) .
\end{align}
Due to the exponential form, the multiple emission case naturally factorizes into products of single emissions. Therefore, non-Abelian exponentiation is manifestly preserved by this regulator.
An important feature of the exponential regulator is that it leads to enormous simplification in perturbative calculations, as demonstrated by the calculation of TMD soft functions at N$^3$LO in Ref.~\cite{Li:2016ctv}.

The exponential regulator also admits simple operator definitions for the TMD functions. For example, the quark TMD soft function is defined as
\begin{equation}
\mathcal{S}^{\text{bare}}_{q\bar{q}}(b_\perp, \nu) \equiv \frac{1}{N_c} \lim_{\tau \to 0}  \Tr 
\braket{0 | [S^\dagger_{\bar{n}} S_n] (-ib_0\tau,-ib_0\tau,b_\perp) \, [S^\dagger_n S_{\bar{n}}] (0) | 0} \Big|_{\tau \equiv 1/\nu} \, ,
\label{eq:TMDsoft}
\end{equation}
where the rapidity regularization procedure is understood as keeping non-vanishing terms in the limit of $\tau \to 0$ (including the $\log\tau$ terms which are the manifestation of rapidity divergences), and then identify the rapidity scale as $\nu = 1/\tau$. No subtraction is needed and the rapidity divergence are now renormalized. The remaining results depend on logarithms of the rapidity scale $\nu$.

Similarly, the exponentially regularized quark TMDPDF and TMDFF are defined as
\begin{multline}
\mathcal{B}^{\text{bare}}_{q/N_1}(x,b_\perp,\nu) \equiv \frac{1}{\mathcal{S}_{\text{0b}}} \lim_{\tau \to 0} \int \frac{db_-}{4\pi} \, e^{-i x b_- P_{1+}/2}
\\
\times \braket{N_1(P_1) | \bar{\chi}_n(-ib_0\tau,b_--ib_0\tau,b_\perp) \frac{\slashed{\bar{n}}}{2} \chi_n(0) | N_1(P_1)} \Big|_{\tau \equiv 1/\nu} \, ,
\label{eq:TMDPDF}
\end{multline}
and
\begin{multline}
\mathcal{D}^{\text{bare}}_{N_2/q}(z,b_\perp,\nu) \equiv \frac{1}{\mathcal{S}_{\text{0b}}} \lim_{\tau \to 0} \sum_X \frac{1}{z} \int \frac{db_+}{4\pi} \, e^{i b_+ P_{2-} / (2z)}
\\
\times \Tr \braket{0 | \frac{\slashed{n}}{2} \chi_{\bar{n}}(b_+-ib_0\tau,-ib_0\tau,b_\perp) | N_2(P_2),X} \braket{N_2(P_2),X | \bar{\chi}_{\bar{n}}(0) | 0} \Big|_{\tau \equiv 1/\nu} \, .
\label{eq:TMDFF}
\end{multline}
Note that for both the TMDPDF and TMDFF, we need to perform a zero-bin subtraction to avoid double-counting between the collinear sectors and the soft sector. The zero-bin soft function is the same as the TMD soft function
\begin{equation}
\mathcal{S}_{\text{0b}}(b_\perp,\nu) = \mathcal{S}^{\text{bare}}_{q\bar{q}}(b_\perp,\nu) \, .
\end{equation}
Having operator definitions Eqs.~\eqref{eq:TMDsoft}, \eqref{eq:TMDPDF} and \eqref{eq:TMDFF} for the TMD functions could be advantageous for studying non-perturbative aspects of TMD physics. In this work we focus on the perturbative part of the TMDPDF and TMDFF.

\subsection{Renormalization and perturbative matching}

For large impact parameter $b_T \sim 1/\Lambda_{\text{QCD}}$, the TMDPDFs and TMDFFs are dominated by long distance contributions and are genuine non-perturbative objects. In this work, we are interested in the semi-perturbative region $b_T \ll 1/\Lambda_{\text{QCD}}$. In this region the TMDPDF admits an operator product expansion
\begin{align}
\mathcal{B}^{\text{bare}}_{q/N}(x,b_\perp,\nu) = \sum_i  \int_x^1 \frac{d\xi}{\xi} \, \mathcal{I}^{\text{bare}}_{qi}(\xi,b_\perp,\nu) \, \phi_{i/N}(x/\xi) + \mathcal{O}(b_T^2\Lambda^2_{\text{QCD}}) \, ,
\label{eq:TMDPDFOPE}
\end{align}
where $\phi_{i/N}$ is the (renormalized) collinear PDF of parton $i$, and $\mathcal{I}_{qi}$ is a perturbatively calculable matching coefficient function describing the splitting of the parton $i$ into the quark $q$. Similarly, the TMDFF can also be factorized as
\begin{align}
\mathcal{F}^{\text{bare}}_{N/q}(z,b_\perp/z,\nu)
&=z^{2-2\e} \mathcal{D}^{\text{bare}}_{N/q}(z,b_\perp,\nu) \nonumber \\
& = \sum_i \int_z^1 \frac{d\xi}{\xi} \, d_{N/i}(z/\xi) \, \mathcal{C}^{\text{bare}}_{iq}(\xi,b_\perp/\xi,\nu) + \mathcal{O}(b_T^2\Lambda^2_{\text{QCD}}) \, .
\label{eq:TMDFFOPE}
\end{align}
with perturbatively calculable coefficient functions $\mathcal{C}_{iq}$ describing the fragmentation of the quark $q$ into the parton $i$.

The functions $\mathcal{I}_{qi}$ and $\mathcal{C}_{iq}$ will be the main objects we are going to study in this work. As indicated by the superscript ``bare'' in Eqs.~\eqref{eq:TMDPDFOPE} and \eqref{eq:TMDFFOPE}, there are UV divergences which require renormalization. For the TMDPDF, we define the renormalization factor according to
\begin{align}
\mathcal{B}^{\text{bare}}_{q/N}(x,b_\perp,\nu) &= Z^B_q(b_\perp,\mu,\nu) \, \mathcal{B}_{q/N}(x,b_\perp,\mu,\nu) \nonumber
\\
&= Z^B_q(b_\perp,\mu,\nu) \sum_i \mathcal{I}_{qi}(x,b_\perp,\mu,\nu) \otimes \phi_{i/N}(x,\mu) + \mathcal{O}(b_T^2\Lambda^2_{\text{QCD}}) \, ,
\end{align}
where we have used $\otimes$ to denote the convolution in Eq.~\eqref{eq:TMDPDFOPE}. The matching coefficients $\mathcal{I}_{qi}$ do not depend on the external state $N$, and can therefore be calculated with $N$ replaced by a partonic state $j=q$ or $g$. We can then extract $\mathcal{I}_{qi}$ by calculating $\mathcal{B}^{\text{bare}}_{q/j}$, performing the renormalization and subtracting the partonic collinear PDFs $\phi_{i/j}$. Up to the NNLO, the partonic collinear PDFs are given by
\begin{align}
\label{eq:phic}
\phi_{i/j}(x,\mu) &= \delta_{ij} \delta(1-x) - \alsmupi \frac{P_{ij}^{\zero}(x)}{\e} \nn
\\
&+ \left( \alsmupi \right)^2
\left[ \frac{1}{2\e^2} \left( \sum_{k} P_{ik}^\zero(x) \otimes P_{kj}^\zero(x) + \beta_0 P_{ij}^\zero(x) \right) - \frac{P_{ij}^\one(x)}{2\e} \right] ,
\end{align}
where $P_{ij}^{(0)}$ is the LO splitting kernel and $\beta_0$ is the LO beta function.

After renormalization, the TMDPDF obeys a renormalization group equation (RGE)
\begin{align}
\frac{\df}{\df\ln\mu} \mathcal{B}_{q/N}(x,b_\perp,\mu,\nu) = 2 \left[ \Gcusp(\alsmu) \ln\frac{\nu}{x P_{1+}} + \gamma^B(\alsmu)  \right] \mathcal{B}_{q/N}(x,b_\perp,\mu,\nu) \, ,
\end{align}
where $\Gcusp$ is the usual cusp anomalous dimension and $\gamma^B$ is the non-cusp anomalous dimension for the TMDPDF, whose perturbative expansions are collected in the Appendix. From the above equation and the famous DGLAP equation
\begin{align}
\frac{\df}{\df\ln\mu} \phi_{i/N}(x,\mu) = 2 \sum_{j} P_{ij}(x,\alsmu) \otimes \phi_{j/N}(x,\mu) \, ,
\end{align}
one can deduce the RGEs for the coefficient functions as
\begin{multline}
\frac{\df}{\df \ln\mu} \cI_{qi}(x,b_\perp,\mu,\nu) = 2 \left[ \Gcusp(\alsmu) \ln\frac{\nu}{xP_{1+}} + \gamma^B(\alsmu) \right] \cI_{qi}(x,b_\perp,\mu,\nu)
\\
- 2 \sum_j \cI_{qj}(x,b_\perp,\mu,\nu) \otimes P_{ji}(x,\alsmu) \, .
\label{eq:Imu}
\end{multline}
Besides the normal RGE, the TMDPDF and the coefficient functions also satisfy the rapidity evolution equation~\cite{Chiu:2012ir}
\begin{equation}
\frac{\df}{\df\ln\nu} \cI_{qi}(x,b_\perp,\mu,\nu) = -2 \left[ \int_{\mu}^{b_0/b_T} \frac{\df\bar{\mu}}{\bar{\mu}} \Gcusp(\alpha_s(\bar{\mu})) + \gamma^R(\als(b_0/b_T)) \right] \cI_{qi}(x,b_\perp,\mu,\nu) \, .
\label{eq:Inu}
\end{equation}
The rapidity anomalous dimension $\gamma^R$ is known to three loops in QCD~\cite{Li:2016ctv,Vladimirov:2016dll}. For our purpose, we need the first two orders which are given by\footnote{Note that the convention here differ by a factor of 2 from Ref.~\cite{Li:2016ctv}.}
\begin{align}
\gamma^R_0 &= 0 \, , \nn
\\
\gamma^R_1 &= C_F \left[ C_A \left( -\frac{404}{27} + 14\zeta_3 \right) + T_FN_f \frac{112}{27} \right] ,
\label{eq:gammaR}
\end{align}

The renormalization equations \eqref{eq:Imu} and \eqref{eq:Inu} can be used to determine all the renormalization and rapidity scale dependent terms for the coefficient functions in perturbation theory.
Throughout this paper, we organize perturbative expansions of various functions in powers of $\alpha_s/(4\pi)$. For example
\begin{align}
\cI_{qi}(x,b_\perp,\mu,\nu) = \sum_{n=0} \left( \alsmupi \right)^n \cI_{qi}^{(n)}(x,b_\perp,L_Q) \, .
\end{align}
Here and below we introduce two logarithms
\begin{equation}
\label{eq:Ldef}
\Lp = \ln\frac{b_T^2\mu^2}{b_0^2} \, , \quad L_Q = 2\ln\frac{xP_{1+}}{\nu} \, .
\end{equation}
Up to $\Ord(\alpha_s^2)$, we then have
\begin{align}
\cI^\zero_{qi}(x,b_\perp,L_Q) &= \delta_{qi} \delta(1-x) \, ,\nn
\\
\cI^\one_{qi}(x,b_\perp,L_Q) &= \left( - \frac{\Gcusp_0}{2} \Lp L_Q + \gamma_0^B \Lp + \gamma_0^R L_Q \right) \delta_{qi} \delta(1-x) - P_{qi}^\zero(x) \Lp + I_{qi}^\one(x) \, , \nn
\\
\cI^\two_{qi}(x,b_\perp,L_Q) &= \bigg[ \frac{1}{8} \left( -\Gcusp_0 L_Q + 2\gamma^B_0 \right) \left( -\Gcusp_0 L_Q + 2\gamma^B_0 + 2\beta_0 \right) \Lp^2 \nn
\\
&\hspace{1em} + \left( -\frac{\Gcusp_1}{2} L_Q + \gamma^B_1 + (-\Gcusp_0 L_Q + 2\gamma^B_0 + 2\beta_0) \frac{\gamma_0^R}{2} L_Q \right) \Lp \nn
\\
&\hspace{1em} + \frac{(\gamma_0^R)^2}{2} L_Q^2 + \gamma_1^R L_Q \bigg] \, \delta_{qi} \delta(1-x) \nn
\\
&+ \bigg( \frac{1}{2} \sum_j P^\zero_{qj}(x) \otimes P^\zero_{ji}(x) + \frac{P^\zero_{qi}(x)}{2} (\Gcusp_0 L_Q - 2\gamma_0^B - \beta_0) \bigg) \Lp^2 \nn
\\
&+ \bigg[ -P^\one_{qi}(x) - P^\zero_{qi}(x) \gamma_0^R L_Q - \sum_j I^\one_{qj}(x) \otimes P^\zero_{ji}(x) \nn
\\
&\hspace{1em} + \left( -\frac{\Gcusp_0}{2} L_Q + \gamma_0^B + \beta_0 \right) I^\one_{qi}(x) \bigg] \Lp + \gamma_0^R L_Q I^\one_{qi}(x) + I^\two_{qi}(x) \, .
\label{eq:RGexpI}
\end{align}

Similarly for the TMDFF, the UV renormalization is given by
\begin{align}
\mathcal{F}^{\text{bare}}_{N/q}(z,b_\perp/z,\nu) &= Z^B_q(b_\perp,\mu,\nu) \, \mathcal{F}_{N/q}(z,b_\perp/z,\mu,\nu) \nonumber
\\
&= Z^B_q(b_\perp,\mu,\nu) \sum_i  d_{N/i}(z,\mu)  \otimes \mathcal{C}_{iq}(z,b_\perp/z,\mu,\nu) + \mathcal{O}(b_T^2\Lambda^2_{\text{QCD}}) \, .
\end{align}
Note that the procedure of renormalization and matching is easier to be done with the parton frame $\mathcal{F}_{N/q}$ instead of the hadron frame $\mathcal{D}_{N/q}$ (used in \cite{Echevarria:2015usa}). To extract the coefficient functions $\cC_{iq}$, we calculate the bare TMDFFs with external parton states, and the partonic collinear FFs up to NNLO are given by
\begin{align}
d_{i/j}(z,\mu) &= \delta_{ij} \delta(1-z) - \alsmupi \frac{P_{ij}^{T\zero}(z)}{\e} \nn
\\
&+ \left( \alsmupi \right)^2
\left[ \frac{1}{2\e^2} \left( \sum_{k} P_{ik}^{T \zero}(z) \otimes P_{kj}^{T \zero}(z) + \beta_0 P_{ij}^{T \zero}(z) \right) - \frac{P_{ij}^{T \one}(z)}{2\e} \right] ,
\end{align}
where $P^T_{ij}(z)$ are the time-like splitting kernels which will be presented in the Appendix. The renormalized $\mathcal{C}_{iq}$ functions satisfy the evolution equations
\begin{multline}
\frac{\df}{\df \ln\mu} \cC_{iq}(z,b_\perp/z,\mu,\nu) = 2 \left[ \Gcusp(\alsmu) \ln\frac{z\nu}{P_{2-}} + \gamma^B(\alsmu) \right] \cC_{iq}(z,b_\perp/z,\mu,\nu)
\\
- 2 \sum_j P^T_{ij}(z,\alsmu) \otimes \cC_{jq}(z,b_\perp/z,\mu,\nu) \, ,
\label{eq:Cmu}
\end{multline}
and
\begin{equation}
\frac{\df}{\df\ln\nu} \cC_{iq}(z,b_\perp/z,\mu,\nu) = -2 \left[ \int_{\mu}^{b_0/b_T} \frac{\df\bar{\mu}}{\bar{\mu}} \Gcusp(\alpha_s(\bar{\mu})) + \gamma^R(\als(b_0/b_T)) \right] \cC_{iq}(z,b_\perp/z,\mu,\nu) \, .
\label{eq:Cnu}
\end{equation}

At this point, it is worth noting that the product of the TMDPDF, TMDFF and the TMD soft function is independent on the rapidity scale $\nu$ as expected, namely
\begin{equation}
\frac{\df}{\df\ln\nu} \big[ \cB_{q/i}(x,b_\perp,\mu,\nu) \, \cD_{j/q}(z,b_\perp,\mu,\nu) \, \cS_{q\bar{q}}(b_\perp,\mu,\nu) \big] = 0 \, ,
\end{equation}
where we have used
\begin{equation}
\frac{d}{d\ln\nu} \mathcal{S}_{q\bar{q}}(b_\perp,\mu,\nu) = 4 \left[ \int_\mu^{b_0/b_T} \frac{d\bar{\mu}}{\bar{\mu}} \, \Gcusp(\als(\bar{\mu})) + \gamma^R(\als(b_0/b_T)) \right] \mathcal{S}_{q\bar{q}}(b_\perp,\mu,\nu) \, .
\label{eq:Snu}
\end{equation}
It can also be shown that the $\mu$-dependence of this product is cancelled by that of the hard function (which does not know about the rapidity divergences), such that the physical observables are independent of the renormalization scale. To see that we recall the RGEs of the hard and soft functions
\begin{align}
\frac{d}{d\ln\mu} H_{q\bar{q}}(Q^2,\mu) &= 2 \left[ \Gcusp(\alsmu) \ln\frac{Q^2}{\mu^2} + 2\gamma^H(\alsmu) \right] H_{q\bar{q}}(Q^2,\mu) \, , \nn
\\
\frac{d}{d\ln\mu} \mathcal{S}_{q\bar{q}}(b_\perp,\mu,\nu) &= 2 \left[ \Gcusp(\alsmu) \ln\frac{\mu^2}{\nu^2} - 2\gamma^S(\alsmu) \right] \mathcal{S}_{q\bar{q}}(b_\perp,\mu,\nu) \, .
\end{align}
We note that the cancellation happens since $\gamma^B + \gamma^H - \gamma^S = 0$ and
\begin{equation}
\ln\frac{\nu}{xP_{1+}} + \ln\frac{z\nu}{P_{2-}} + \ln\frac{\mu^2}{\nu^2} = \ln\frac{\mu^2}{Q^2} \, .
\end{equation}

By using the evolution equations, we can derive the scale-dependent part of $\cC_{iq}$. Up to the NNLO we have
\begin{align}
\cC^\zero_{iq}(z,b_\perp/z,L_Q) &= \delta_{iq} \delta(1-z) \, ,\nn
\\
\cC^\one_{iq}(z,b_\perp/z,L_Q) &= \left( - \frac{\Gcusp_0}{2} \Lp L_Q + \gamma_0^B \Lp + \gamma_0^R L_Q \right) \delta_{iq} \delta(1-z) - P_{iq}^{T\zero}(z) \Lp + C_{iq}^\one(z) \, , \nn
\\
\cC^\two_{iq}(z,b_\perp/z,L_Q) &= \bigg[ \frac{1}{8} \left( -\Gcusp_0 L_Q + 2\gamma^B_0 \right) \left( -\Gcusp_0 L_Q + 2\gamma^B_0 + 2\beta_0 \right) \Lp^2 \nn
\\
&\hspace{1em} + \left( -\frac{\Gcusp_1}{2} L_Q + \gamma^B_1 + (-\Gcusp_0 L_Q + 2\gamma^B_0 + 2\beta_0) \frac{\gamma_0^R}{2} L_Q \right) \Lp \nn
\\
&\hspace{1em} + \frac{(\gamma_0^R)^2}{2} L_Q^2 + \gamma_1^R L_Q \bigg] \, \delta_{iq} \delta(1-z) \nn
\\
&+ \bigg( \frac{1}{2} \sum_j P^{T\zero}_{ij}(z) \otimes P^{T\zero}_{jq}(z) + \frac{P^{T\zero}_{iq}(z)}{2} (\Gcusp_0 L_Q - 2\gamma_0^B - \beta_0) \bigg) \Lp^2 \nn
\\
&+ \bigg[ -P^{T\one}_{iq}(z) - P^{T\zero}_{iq}(z) \gamma_0^R L_Q - \sum_j P^{T\zero}_{ij}(z) \otimes C^\one_{jq}(z) \nn
\\
&\hspace{1em} + \left( -\frac{\Gcusp_0}{2} L_Q + \gamma_0^B + \beta_0 \right) C^\one_{iq}(z) \bigg] \Lp + \gamma_0^R L_Q C^\one_{iq}(z) + C^\two_{iq}(z) \, .
\label{eq:RGexpC}
\end{align}
Note that here we have used the same symbol $L_Q$ as in Eq.~\eqref{eq:Ldef} to denote a different meaning:
\begin{equation}
L_Q = 2 \ln \frac{P_{2-}}{z \nu} \, ,
\end{equation}
which can be regarded as the crossing $P_{1+} \to P_{2-}$ and $x \to 1/z$.

\subsection{Rapidity renormalization group and re-factorization}

We now use the rapidity evolution equations of the TMDPDF, TMDFF and TMD soft function to derive the re-factorization formula \eqref{eq:refac}. From the perturbative matching coefficients, it is evident that for the TMDPDF and TMDFF, the natural rapidity scale is $\nu \sim xP_{1+} \sim P_{2-}/z \sim Q$, while for the TMD soft function the natural choice is $\nu \sim b_0/b_T \sim q_T$. In order to reconcile these different choices, we may use the rapidity RGE \eqref{eq:Snu} for the TMD soft function to evolve it from $\nu = b_0/b_T$ to $\nu=Q$. The result is
\begin{equation}
\mathcal{S}_{q\bar{q}}(b_\perp,\mu,\nu=Q) = \mathcal{S}_{q\bar{q}}(b_\perp,\mu,\nu=b_0/b_T) \left( \frac{b_T^2Q^2}{b_0^2} \right)^{-F_{q\bar{q}}(L_\perp,\alpha_s(\mu))} \, ,
\end{equation}
where
\begin{align}
F_{q\bar{q}}(L_\perp,\alpha_s(\mu)) &\equiv -2 \left[ \int_\mu^{b_0/b_T} \frac{d\bar{\mu}}{\bar{\mu}} \, \Gcusp(\als(\bar{\mu})) + \gamma^R(\als(b_0/b_T)) \right] \nn
\\
&= \alsmupi \Gcusp_0 \Lp + \left( \alsmupi \right)^2 \left( \frac{\beta_0\Gcusp_0}{2} \Lp^2 + \Gcusp_1 \Lp - 2 \gamma_1^R \right) + \Ord(\als^3) \, ,
\end{align}
where we have used $\gamma_0^R=0$.
The ``genuine'' quark TMDPDF and TMDFF which are free from rapidity divergences and are independent of the hard scale can then be defined as
\begin{align}
B_{q/N_1}(x,b_\perp,\mu) &\equiv \cB_{q/N_1}(x,b_\perp,\mu,\nu=Q) \, \sqrt{\cS_{q\bar{q}}(b_\perp, \mu,\nu=b_0/b_T)} \,, \nn
\\
D_{N_2/q}(z,b_\perp,\mu) &\equiv \cD_{N_2/q}(z,b_\perp,\mu,\nu=Q) \, \sqrt{\cS_{q\bar{q}}(b_\perp, \mu,\nu=b_0/b_T)} \, .
\end{align}
These are essentially the functions appearing in the re-factorization formula \eqref{eq:refac}.

%
%

\section{Quark TMDPDF with the exponential regulator}
\label{sec:tmdpdf}

In this section, we calculate the perturbative matching coefficients of the quark TMDPDF at NLO and NNLO using the exponential regulator. While these results are known to order $\epsilon^0$ in the literature \cite{Gehrmann:2012ze,Gehrmann:2014yya,Echevarria:2016scs}, we are able to obtain higher order terms in $\epsilon$. The calculation with the exponential regulator is also much simpler and more systematic, which makes it possible to be extended to N$^3$LO.

\subsection{Quark TMDPDF at NLO}

In this subsection, we briefly discuss the NLO results with the exponential regulator. While the calculation is straightforward, it illustrates the basic procedure and some interesting features of the regularization scheme.

\begin{figure}[t!]
\centering
\includegraphics[width=0.8\linewidth]{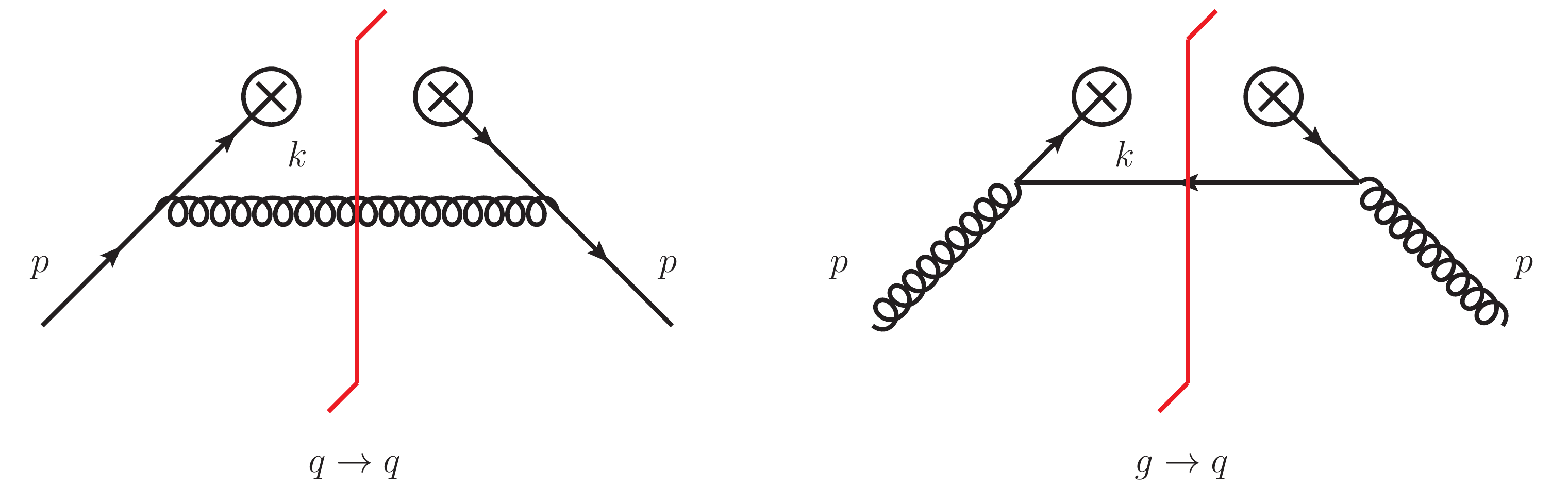}
\caption{\label{fig:NLO}Cut diagrams for the bare coefficient functions at NLO.}
\end{figure}

We begin with the bare TMDPDFs before zero-bin subtraction. According to the definition in Eqs.~\eqref{eq:TMDPDF}, the TMDPDFs at NLO are given by the cut diagrams in Fig.~\ref{fig:NLO}. The result can be written as
\begin{multline}
\alspi \cB_{q/i}^{\one,\text{bare},\text{unsub}}(x,b_\perp,\nu) = \lim_{\tau \to 0} \int \frac{\df^dk}{(2\pi)^d} \, (2\pi) \delta_+(k^2) \, \delta(k_+ - (1-x) p_{+})
\\
\times  \frac{g_s^2 \mu^{2\e} k_+}{\kpsq} \, p_{qi}^\zero(x,\e) \, \exp \left[- \frac{b_0 \tau}{2}(k_+ + k_-) + i \bp \cdot \kp \right] \bigg|_{\tau=1/\nu} \, ,
\end{multline}
where we have changed to the notation that $\kp$ denotes the transverse components of $k_\perp^\mu$, but with Euclidean signature such that
\begin{equation}
\bp \cdot \kp = - b_\perp \cdot k_\perp \, , \quad \kpsq = |\kp|^2 = -k_\perp^2 \, .
\end{equation}
The $d$-dimensional splitting amplitudes are given by
\begin{align}
p_{qq}^\zero (x,\e) &= 2 C_F \left[ \frac{1 + x^2}{1 - x} - \e ( 1 - x ) \right] , \nn
\\
p_{qg}^\zero (x,\e) &= 2 T_F \left[ 1 - \frac{2}{1-\e} x(1-x) \right] \, .
\label{eq:splitting}
\end{align}
Using the delta function for $k_+$ and the on-shell condition, we can
write the exponential regulator as
\begin{equation}
\exp \left[ - \frac{b_0 \tau}{2}(k_+ + k_-) + i \bp \cdot \kp \right]
= \exp \left[ - \frac{b_0 \tau}{2} \left( \frac{\kpsq}{(1-x)p_+} + (1-x) p_+ \right) + i \bp \cdot \kp \right] .
\end{equation}
At this stage we can already drop the second term proportional to $\tau(1-x)$ in the exponent, as it gives no contribution in the limit $\tau \to 0$.
They might be relevant for subleading power corrections~\cite{Ebert:2018gsn}.
The first term involving $\tau/(1-x)$ in the exponent provides the main service of regularizing the rapidity divergences. To see how that happens, we note that the rapidity divergence appears here as a singularity as $x \to 1$ in the $q \to q$ splitting amplitude in Eq.~\eqref{eq:splitting}. The exponential regulator provides a suppression in the $x \to 1$ limit, and turns this singularity into a regularized distribution according to
\begin{align}
\frac{e^{-\tau/(1-x)}}{1-x} = - (\ln \tau + \gamma_E) \, \delta(1-x) + \frac{1}{(1-x)_+} + \cO(\tau) \, .
\label{eq:expreg_dist1}
\end{align}
Applying the above equation to $\cB_{q/q}^{\one,\text{bare},\text{unsub}}$, we find
\begin{multline}
\cB_{q/q}^{\one,\text{bare},\text{unsub}}(x,b_\perp,\nu) = \frac{e^{\e\gamma_E} \mu^{2\e}}{\pi^{1-\e}} C_F \big( 1 + x^2 - \e (1-x)^2 \big) \int \frac{\df^{2-2\e} \kp}{\kpsq} e^{i \bp \cdot \kp}
\\
\times \left[ \left( \ln\frac{\mu^2}{\kpsq} - \ln \frac{\mu^2}{\nu p_+} \right) \delta(1-x) + \frac{1}{(1-x)_+} \right]  + \cO(\tau) \,.
\label{eq:Iqq1}
\end{multline}
The $\kp$ integral can be easily performed with the help of the generating integral
\begin{align}
\frac{e^{\e \gamma_E} \mu^{2\e}}{\pi^{1-\e}} \int \frac{\df^{2-2\e} \kp}{\kp^{2 + 2 \eta}} e^{i \bp \cdot \kp} = e^{-(2 \e + 2 \eta) \gamma_E} \mu^{-2 \e - 2 \eta}
\frac{\Gamma(-\e - \eta)}{\Gamma(1 + \eta)} e^{(\e + \eta) \Lp} \, .
\label{eq:kTint}
\end{align}
In particular, we have
\begin{align}
\frac{1}{\pi^{1-\e}} \int \frac{d^{2-2 \e} \kp}{\kpsq} e^{i \bp \cdot \kp} &= e^{-2 \e \gamma_E} \mu^{-2 \e} \Gamma(-\e) e^{\e \Lp} \, , \nn
\\
\frac{1}{\pi^{1-\e}} \int \frac{d^{2-2 \e} \kp}{\kpsq} e^{i \bp \cdot \kp} \ln\frac{\mu^2}{\kpsq} &= e^{-2 \e \gamma_E} \mu^{-2 \e} \Gamma(-\e) e^{\e\Lp} [ \Lp - \gamma_E - \psi (-\e) ]\,,
\end{align}
where $\psi(x) = \Gamma'(x)/\Gamma(x)$. Applying the above formulas to Eq.~\eqref{eq:Iqq1} then gives $\cB_{q/q}^{\one,\text{bare},\text{unsub}}$ exact in $\e$, and similarly for $\cB_{q/g}^{\one,\text{bare},\text{unsub}}$. It is easy to expand the results to any order in $\e$. In particular, the $\Ord(\e^2)$ terms will be used for the NNLO calculations later, while the $\Ord(\e^4)$ terms are relevant for the calculations at N$^3$LO.

We now need to subtract the collinear PDFs in Eq.~\eqref{eq:phic} to obtain the coefficient functions $\cI^{\one,\text{bare},\text{unsub}}_{qi}$, and then perform the zero-bin subtraction, where the NLO zero-bin contribution is given by
\begin{align}
\mathcal{S}_{\text{0b}}^\one(\Lp,\mu,\nu) &= \mathcal{S}_{q\bar{q}}^{\one,\text{bare}}(\Lp,\mu,\nu) 
= C_F \left( \frac{4}{\e^2} - \frac{4}{\e} L_\nu -2 \Lp^2 - 4 \Lp L_\nu - \frac{\pi^2}{3} \right) ,
\end{align}
where $L_\nu=\ln(\nu^2/\mu^2)$.
After the subtraction, we find up to $\Ord(\e^0)$ 
\begin{align}
\cI_{qq}^{\one,\text{bare}}(x,b_\perp,L_Q) &= \cI_{qq}^{\one,\text{bare},\text{unsub}}(x,b_\perp,\mu,\nu) - S_{\text{0b}}^\one(\Lp,\mu,\nu) \, \delta(1-x) \nn
\\
&= C_F \left( \frac{1}{\e} + \Lp \right) ( -2 L_Q  + 3 ) \, \delta(1-x)  - \Lp P_{qq}^\zero(x) + 2 C_F ( 1 - x ) \,,
\nn\\
\cI_{qg}^{\one,\text{bare}}(x,b_\perp,L_Q) &= \cI_{qg}^{\one,\text{bare},\text{unsub}}(x,b_\perp,\mu,\nu) = 2T_F-(1+\Lp)P_{qg}^\zero(x)\, ,
\end{align}
where $L_Q = 2\ln(xp_+/\nu)$, and
\begin{align}
P_{qq}^\zero(x) &= C_F \left[3 \delta(1-x) + \frac{2(1 + x^2)}{(1-x)_+} \right] , \nn
\\
P_{qg}^\zero(x) &= 2 T_F \left[ (1-x)^2 + x^2 \right] .
\end{align}
We now renormalize the UV divergences in the $\overline{\text{MS}}$ scheme with the renormalization factor
\begin{equation}
Z_q^B(b_\perp,\mu,\nu) = 1 +\frac{\alpha_s}{4\pi}\frac{\Gcusp_0 L_Q-2\gamma^B_0}{-2\e} + \Ord(\als^2) \, ,
\end{equation}
and find
\begin{align}
\cI_{qq}^{\one}(x,b_\perp,L_Q) &= C_F \, \Lp \, ( -2 L_Q  + 3 ) \, \delta(1-x)  - \Lp P_{qq}^\zero(x) + 2 C_F ( 1 - x )  \,, \nn
\\
\cI_{qg}^{\one}(x,b_\perp,L_Q) &= - \Lp P_{qg}^\zero(x) + 4 T_F x (1 - x) \,.
\end{align}
Comparing the above form with Eq.~\eqref{eq:RGexpI}, we can extract the (renormalization and rapidity) scale-independent part of the NLO coefficients
\begin{align}
I_{qq}^\one(x) &= 2 C_F (1 - x) \,, \nn
\\
I_{qg}^\one(x) &= 4 T_F x (1 - x) \,.
\end{align}
Remarkably, in the exponential regularization scheme, the scale independent coefficients are regular in the soft limit $x \to 1$. As will be explicitly shown below, at NNLO there are $1/(1-x)_+$ distributions in the $\mu$-independent part, but these terms are governed by the rapidity anomalous dimension and depend on the rapidity scale $\nu$. In general this is true even at high orders in perturbation theory~\cite{Echevarria:2016scs,Lustermans:2016nvk}.  

\subsection{Quark TMDPDF at NNLO}

We now turn to the NNLO calculations. As before, we begin with the bare TMD coefficient functions before zero-bin subtraction. At NNLO, diagrammatically there are two kinds of contributions. One is the interference of the LO amplitude with the diagrams containing one loop and one real emission, i.e., the so-called real-virtual (RV) contribution. The other is the square of the diagrams with two real emissions, i.e., the so-called double real contribution~(RR). We will discuss these two contributions one-by-one in the following.
  
\subsubsection{The real-virtual contribution}

\begin{figure}[t!]
\centering
\includegraphics[width=0.95\linewidth]{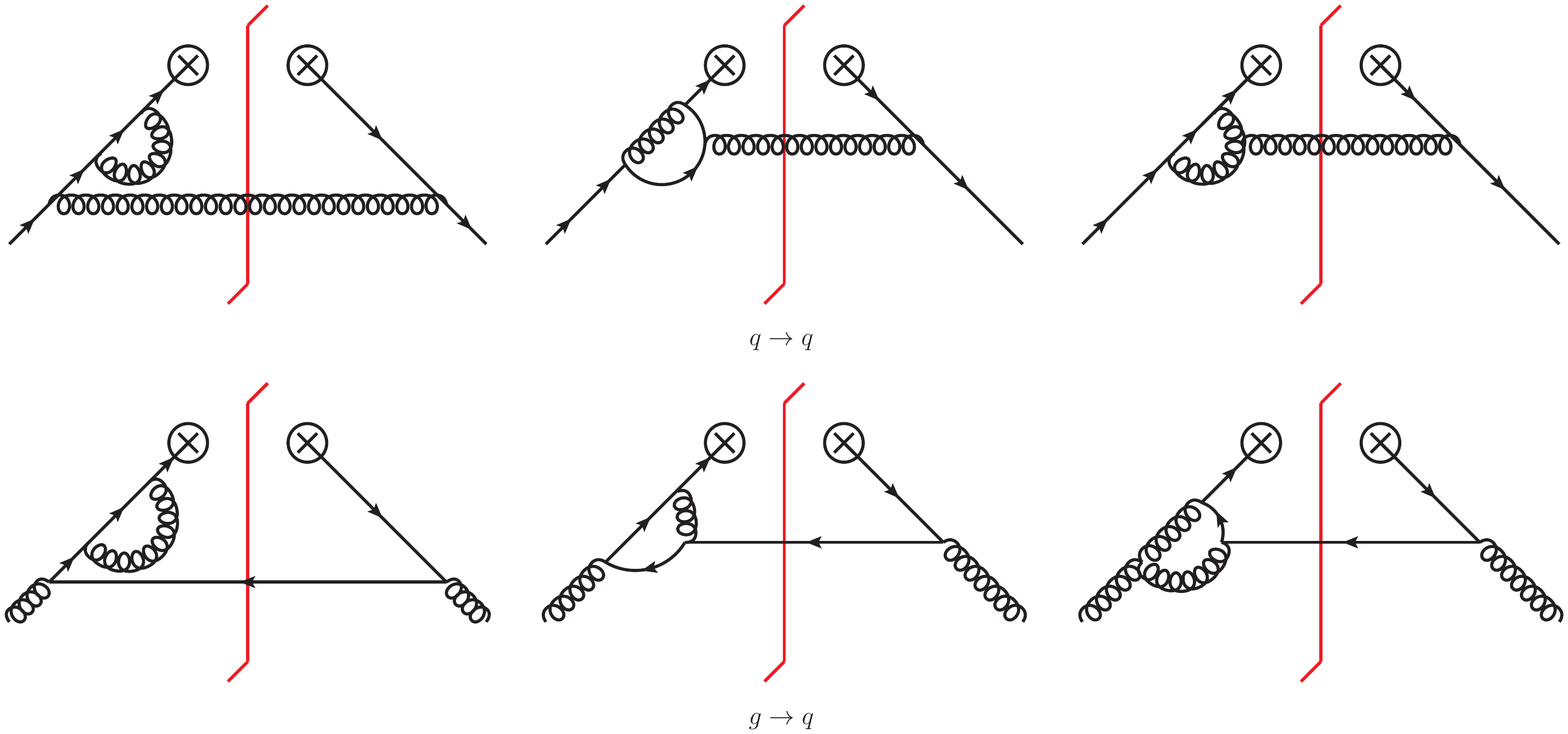}
\vspace{-12ex}
\caption{\label{fig:NNLOVR}Cut diagrams for the real-virtual contribution.}
\end{figure}

\begin{figure}[t!]
\centering
\includegraphics[width=0.85\linewidth]{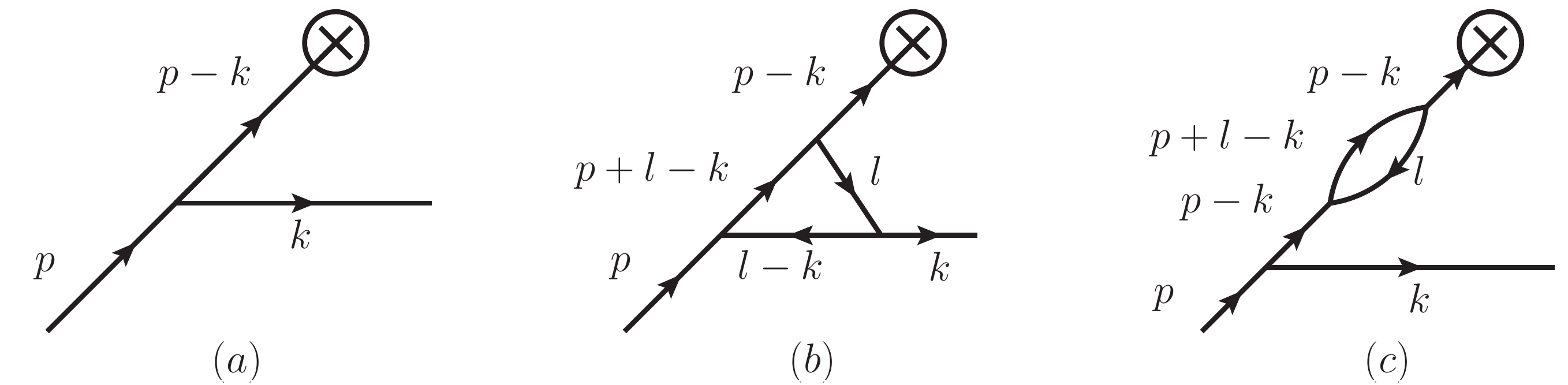}
\caption{\label{fig:VRMT}Topologies for the real-virtual contribution.}
\end{figure}

We adopt the light-cone gauge $n \cdot A = 0$ where the relevant cut diagrams for the real-virtual contribution are depicted in Figure~\ref{fig:NNLOVR}. Note that with both the exponential regulator and the analytic regulator used in \cite{Gehrmann:2012ze,Gehrmann:2014yya}, the loop integral does not need to be regularized. Therefore the treatments of the loop amplitude are rather similar. After performing the Dirac algebras and partial fractioning, there remain two classes of scalar integrals as shown in Fig.~\ref{fig:VRMT}, which are given by
\begin{align}
I^{\text{RV}}_1(a_1,a_2,a_3,a_4) &= \int \frac{d^d l}{(2\pi)^d} \left[ -l^2 \right]^{-a_1} \left[ -(l+q)^2 \right]^{-a_2} \left[ -(l+p)^2 \right]^{-a_3} \left[ \bar{n} \cdot l \right]^{-a_4} \, , \nonumber
\\
I^{\text{RV}}_2(a_1,a_2,a_3,a_4) &= \int \frac{d^d l}{(2\pi)^d} \left[ -l^2 \right]^{-a_1} \left[ -(l+q)^2 \right]^{-a_2} \left[ -(l-k)^2 \right]^{-a_3} \left[ \bar{n} \cdot l \right]^{-a_4} \, ,
\end{align}
where $q=p-k$ and we make the $+i\epsilon$ prescription for all propagators implicit. The results of these integrals have already been given in \cite{Gehrmann:2014yya} and we do not repeat them here. After the loop integration, the results are functions of $x = \bar{n} \cdot q / \bar{n} \cdot p$ and $k_T^2$. The remaining integral over $\kpsq$ can be carried out in the same way as the NLO calculation using Eq.~\eqref{eq:kTint}.

\subsubsection{The double real contribution}

\begin{figure}[t!]
\centering
\includegraphics[width=0.95\linewidth]{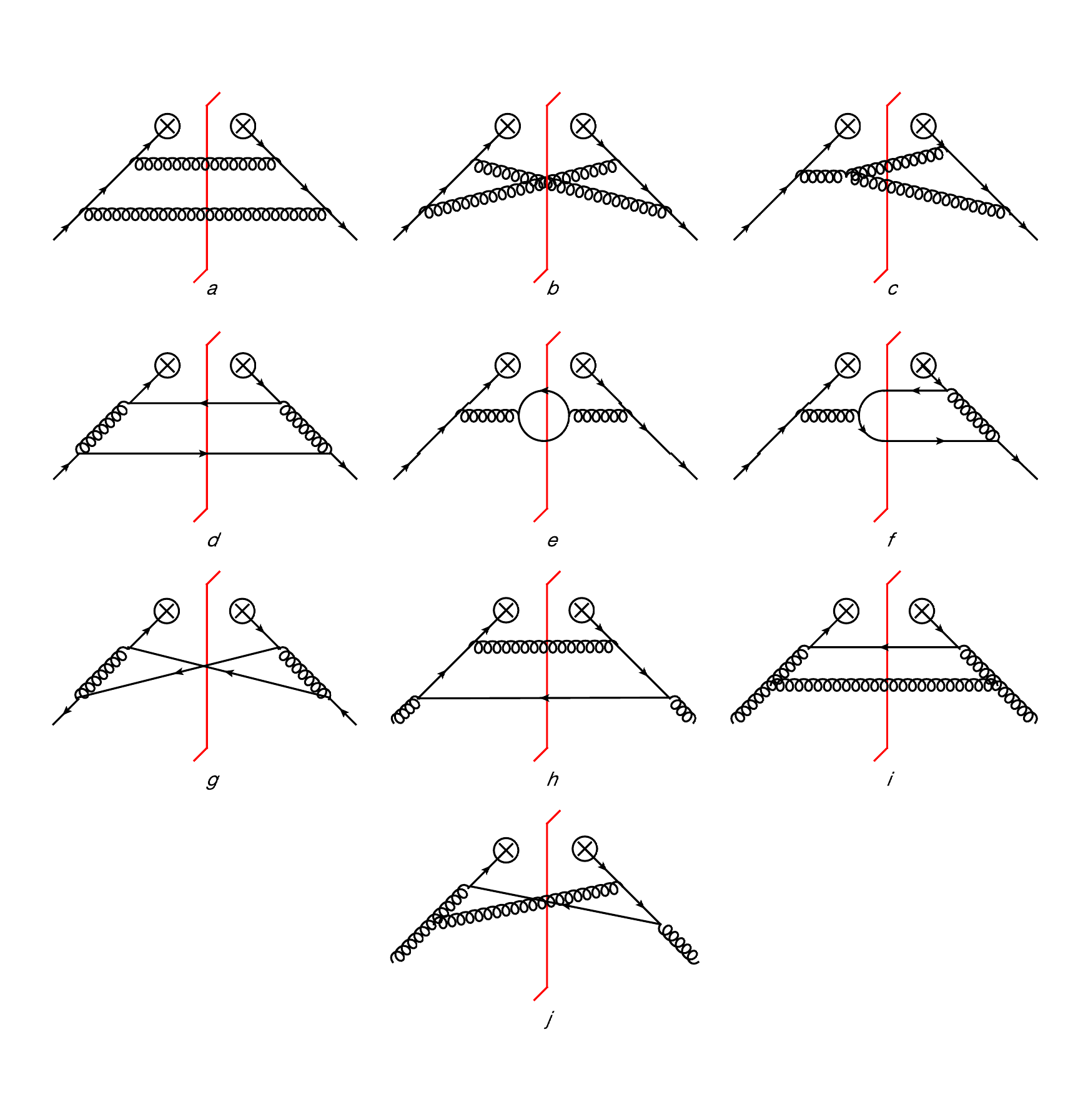}
\vspace{-9ex}
\caption{\label{fig:NNLORR}Cut diagrams for the double-real contribution.}
\end{figure}

\begin{figure}[t!]
\centering
\includegraphics[width=0.85\linewidth]{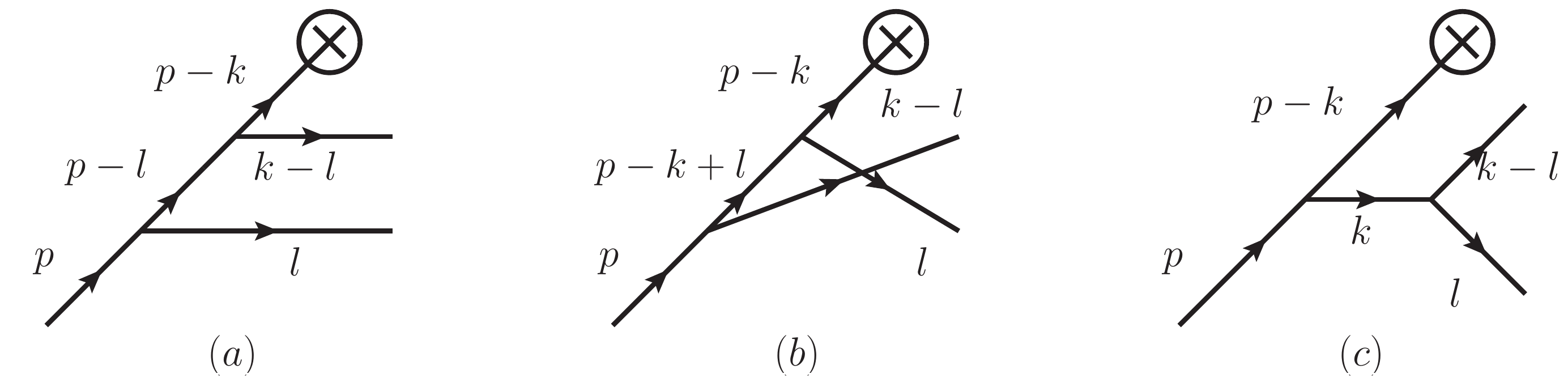}
\caption{\label{fig:RRMT}Topologies for the double-real contribution.}
\end{figure}

At NNLO, the double real contribution is the most troublesome one to calculate. We will show that with the exponential regulator, we can apply many modern techniques for loop integrals. It is therefore possible to extend the calculation method to higher orders. 

We use \texttt{QGRAF}~\cite{Nogueira:1991ex} to generate the relevant Feynman diagrams in the light-cone gauge, which are shown in Fig.~\ref{fig:NNLORR}. We then use \texttt{FORM}~\cite{Vermaseren:2000nd} to manipulate the squared amplitudes, and write them as integrals over the two cut momenta which we denote as $k_1$ and $k_2$. We now need to apply the exponential regulator, and the integral measure then becomes
\begin{equation}
\int \frac{d^dk_1}{(2\pi)^d} \frac{d^dk_2}{(2\pi)^d} \, (2\pi) \delta_+(k_1^2) \, (2\pi) \delta_+(k_2^2) \, \exp \left( -b_0 \tau (k_1^0+k_2^0) \right) .
\end{equation}
It is useful to introduce an identity~\cite{Zhu:2014fma} 
\begin{equation}
\int d^dk \, \delta^{(d)}(k - k_1 - k_2) = 1 \, ,
\end{equation}
and rewrite the integral measure as 
\begin{equation}
\int d^dk \, e^{-b_0 \tau k^0}
\int \frac{d^dk_1}{(2\pi)^d} \frac{d^dk_2}{(2\pi)^d} \, (2\pi) \delta_+(k_1^2) \, (2\pi) \delta_+(k_2^2)
\, \delta^{(d)}(k - k_1 - k_2) \, .
\end{equation}
Now the integration over $k_1$ and $k_2$ does not produce rapidity divergences and can be performed with usual techniques. Note that this fact holds also beyond NNLO where more than two cut momenta are present, due to the exponential form of the regulator.

We can now use the delta function to integrate over $k_2$, and rename $k_1$ as $l$. The double real contribution can then be written in the form
\begin{multline}
\left( \alsmupi \right)^2 \mathcal{B}_{qi}^{(2),\text{RR}}(x,b_\perp,\nu) =  \lim_{\tau \rightarrow 0} \int \frac{d^{d}k}{(2\pi)^{d-1}} \, \exp \left[ - \frac{b_0 \tau}{2} (k_+ + k_-) + i \bp \cdot \kp \right]
\\
\times \delta( k_+ - (1-z)p_+ )
\int \frac{d^{d}l}{(2\pi)^{d-1}} \, \delta_{+}(l^2) \, \delta_{+}((k-l)^2) \, \mathcal{M}_{qi}(p,k,l,\bar{n}) \, ,
\label{eq:I2RR}
\end{multline}
where ${\cal M}_{qi}$ is the squared amplitude. We will first integrate over $l$ using the methods of reverse unitarity~\cite{Anastasiou:2002yz}, integration-by-parts (IBP)~\cite{Chetyrkin:1981qh} and differential equations~\cite{Bern:1993kr,Gehrmann:1999as,Henn:2013pwa}. The relevant topologies are given by (the square of) the diagrams shown in Figure~\ref{fig:RRMT}. There are 4 topologies for the $l$-integrals, which are defined by
\begin{align}
  I^{\text{RR}}_1(a_1,a_2,a_3,a_4) &= \int \frac{d^d l}{(2\pi)^{d-1}} \left[ -l^2 \right]_{\rm cut}^{-a_1} \left[ -(k-l)^2 \right]_{\rm cut}^{-a_2} \left[ p\cdot l \right]^{-a_3} \left[ \bar{n} \cdot l \right]^{-a_4} \, , \nonumber
  \\
  I^{\text{RR}}_2(a_1,a_2,a_3,a_4) &= \int \frac{d^d l}{(2\pi)^{d-1}} \left[ -l^2 \right]_{\rm cut}^{-a_1} \left[ -(k-l)^2 \right]_{\rm cut}^{-a_2} \left[ p\cdot(k-l) \right]^{-a_3} \left[ \bar{n} \cdot l \right]^{-a_4} \, , \nonumber
  \\
  I^{\text{RR}}_3(a_1,a_2,a_3,a_4) &= \int \frac{d^d l}{(2\pi)^{d-1}} \left[ -l^2 \right]_{\rm cut}^{-a_1} \left[ -(k-l)^2 \right]_{\rm cut}^{-a_2} \left[ p\cdot l \right]^{-a_3} \left[ \bar{n} \cdot (p-l) \right]^{-a_4} \, , \nonumber
  \\
  I^{\text{RR}}_4(a_1,a_2,a_3,a_4) &= \int \frac{d^d l}{(2\pi)^{d-1}} \left[ -l^2 \right]_{\rm cut}^{-a_1} \left[ -(k-l)^2 \right]_{\rm cut}^{-a_2} \left[ p\cdot(k-l) \right]^{-a_3} \left[ \bar{n} \cdot (p-l) \right]^{-a_4} \, ,
\end{align}
where we use the subscript ``cut'' to label the cut propagators~\cite{Anastasiou:2002yz}.
Integrals in each topology are further reduced to a set of Master Integrals~(MIs) by IBP identities~\cite{Chetyrkin:1981qh}. In this work, we use \texttt{FIRE5}~\cite{Smirnov:2014hma} and \texttt{LiteRed}~\cite{Lee:2012cn} to perform the reduction. In total we have 6 MIs which can be chosen as
\begin{align}
F_1 &= N(\epsilon) \, w_1 \int d^dl \, \delta_+(l^2) \, \delta_+((k-l)^2) \, ,  \nn
\\
F_2 &= N(\epsilon) \, w_2 \int d^dl \, \frac{\delta_+(l^2) \, \delta_+((k-l)^2)}{\bar{n} \cdot (p-l)} \, , \nn
\\
F_3 &= N(\epsilon) \, w_3 \int d^dl \, \frac{\delta_+(l^2) \, \delta_+((k-l)^2)}{\bar{n} \cdot (p-l) \; p \cdot l} \, , \nn
\\
F_4 &= N(\epsilon) \, w_4 \int d^dl \, \frac{\delta_+(l^2) \, \delta_+((k-l)^2)}{\bar{n} \cdot l \; p \cdot l} \, , \nn
\\
F_5 &= N(\epsilon) \, w_5 \int d^dl \, \frac{\delta_+(l^2) \, \delta_+((k-l)^2)}{\bar{n} \cdot (p-l) \; p \cdot (k-l)} \, , \nn
\\ 
F_6 &= N(\epsilon) \, w_6 \int d^dl \, \frac{\delta_+(l^2) \, \delta_+((k-l)^2)}{\bar{n} \cdot l \; p \cdot (k-l)} \, ,
\end{align} 
where the normalization factor
\begin{equation}
N(\epsilon) = \frac{2\Gamma(2-2\epsilon)}{\pi^{1-\epsilon} \, \Gamma(1-\epsilon)} \, .
\end{equation}
We will use the method of differential equations to evaluate these MIs. For that purpose we have introduced the rescale factors $w_i$ to convert the MIs into a canonical basis~\cite{Henn:2013pwa}. They are given by
\begin{align}
\label{eq:wi}
w_1&=(k^2)^\epsilon \,, \nonumber
\\
w_2&=(\bar{n}\cdot p)(1-x)\frac{\epsilon}{1-2\epsilon} \,, \nonumber
\\
w_3&=(k^2)^{1+\epsilon}(\bar{n}\cdot p)\frac{xy-x-y}{2(1-x)(1-y)}\frac{\epsilon}{1-2\epsilon} \,, \nonumber
\\
w_4&=(k^2)^{1+\epsilon}(\bar{n}\cdot p)\frac{\epsilon}{1-2\epsilon}\frac{-1}{4} \,, \nonumber
\\
w_5&=(k^2)^{1+\epsilon}(\bar{n}\cdot p)\frac{-xy+y-1}{2(1-x)(1-y)}\frac{\epsilon}{1-2\epsilon} \,, \nonumber
\\
w_6&=(k^2)^{1+\epsilon}(\bar{n}\cdot p)\frac{\epsilon}{1-2\epsilon}\frac{-1}{4}\frac{y}{1-y} \,,
\end{align}
where the dimensionless variables $x$ and $y$ are defined as
\begin{equation}
1-x = \frac{ \bar{n} \cdot k}{ \bar{n} \cdot p} \, , \quad 1-y = \frac{k^2 \; \bar{n} \cdot p}{2 \, p \cdot k \; \bar{n} \cdot k} \, .
\end{equation}
The factors $\omega_i$ can be easily obtained using an in-house code or the program package \texttt{CANONICA}~\cite{Meyer:2017joq} which implements the algorithm of~\cite{Meyer:2016slj}.

Among all the MIs, $F_1$, $F_2$, $F_4$ and $F_6$ are easy to be evaluated in closed form
\begin{align}
F_1 &=1\,,\nonumber \\
F_2 &=(1-x)\frac{\epsilon}{1-2\epsilon} \, {}_2F_1(1,1-\epsilon,2-2\epsilon,1-x)\,,\nonumber \\
F_4 &=(1-y)^{-\epsilon} \, {}_2F_1(-\epsilon,-\epsilon,1-\epsilon,y)\,,\nonumber \\
F_6 &=y^{-\epsilon} \, {}_2F_1(-\epsilon,-\epsilon,1-\epsilon,1-y)\,,
\end{align}
where ${}_2F_1$ is the hypergeometric function. $F_3$ and $F_5$ depend on both $x$ and $y$ and are more difficult to calculate. We can construct the differential equations of them with respect to $y$
\begin{align}
\label{eq:DE1}
\frac{\partial F_3}{\partial y} &= \epsilon \left[ \left(\frac{1}{1-y}-\frac{1}{y}\right)F_1 + \left(-\frac{1}{1-y}-\frac{1}{y}\right)F_2 + \left(\frac{1}{y}-\frac{1}{1-y}-\frac{2(1-x)}{xy-x-y}\right)F_3 \right],\nonumber
\\
\frac{\partial F_5}{\partial y} &= \epsilon \left[\left(\frac{1}{1-y}-\frac{1}{y}\right)F_1 + \left(\frac{1}{1-y}+\frac{1}{y}\right)F_2 + \left(\frac{1}{y}-\frac{1}{1-y}+\frac{2(1-x)}{xy-x-y}\right)F_5 \right],
\end{align} 
which are in the so-called canonical form~\cite{Henn:2013pwa}. Given their boundary conditions at $y=0$
\begin{align}
F_3(x,y=0) &= x \; {}_2F_1(1,1-\epsilon,1-2\epsilon,1-x)\,,\nonumber \\
F_5(x,y=0) &= {}_2F_1(1,-\epsilon,1-2\epsilon,1-x)\,,
\end{align}
it is easy to solve the differential equations order-by-order in $\e$ in terms of Goncharov multiple polylogarithms (GPLs). We have obtained the solutions up to weight 6, which will be sufficient for a future N$^3$LO calculation.

The next step is then to perform the remaining integration over $k$ in Eq.~\eqref{eq:I2RR}. The $k_+$ integral can be done using the delta function. And the $k_-$ integral can be changed to use the $y$ variable through
\begin{equation}
\int dk_- = \int_0^1 \frac{dy}{y^2} \frac{\kpsq}{(1-x)p_+} \, .
\end{equation}
Note that we now have singularities at $y \to 0$ or $x \to 1$, which are both manifestations of rapidity divergences. These overlapping singularities often make high order perturbative calculations difficult due to the fact that the regularized integrand is often a complicated function of $x$ and $y$. In our scheme, the regularization is provided by
\begin{align}
\exp \left[ - \frac{b_0 \tau}{2}(k^+ + k^-) \right]
=
\exp \left[- \frac{b_0 \tau}{2} \left( \frac{\kpsq}{(1-x) y p_+} + (1-x) p_+ \right) \right] ,
\end{align}
where the $y \to 0$ and $x \to 1$ limits are both exponentially suppressed. To perform the integration over $y$, we expand the above exponential regulator in terms of delta functions and plus-distributions according to Eq.~\eqref{eq:expreg_dist1} and
\begin{multline}
\frac{1}{(1-x)y} \exp \left( -\frac{\tau}{(1-x)y} \right) = \left( \frac{1}{2}(\ln \tau + \gamma_E)^2 + \frac{\pi^2}{12} \right) \delta(1-x) \, \delta(y) + \frac{1}{(1-x)_+} \frac{1}{y_+}
\\
+ \left( \bigg[ \frac{\ln y}{y} \bigg]_+ - \frac{\ln \tau + \gamma_E}{(y)_+} \right) \delta(1-x)
+ \left( \bigg[ \frac{\ln(1-x)}{1-x} \bigg]_+ - \frac{\ln \tau + \gamma_E}{(1-x)_+} \right) \delta(y)
+ \cO(\tau) \,.
\end{multline}
The $y$ integration can now be done using the package \texttt{HyperInt}~\cite{Panzer:2014caa} and the $\kp$ integration can again be evaluated with the help of Eq.~\eqref{eq:kTint}. After the integration, the results can be expressed in terms of Harmonic PolyLogarithms (HPLs)~\cite{Remiddi:1999ew} of the variable $x$. We use the program package \texttt{HPL}~\cite{Maitre:2005uu} to deal with these functions.

\subsubsection{Final results at NNLO}

Combining the real-virtual and double-real contributions, we obtain the bare un-subtracted NNLO TMDPDF. We then perform the zero-bin subtraction to remove double-counting between the collinear and soft sectors, and apply the usual $\alpha_s$ renormalization and operator renormalization $Z^B_q$ to remove the UV divergences. We have reproduced all the renormalization and rapidity scale dependent parts in Eq.~\eqref{eq:RGexpI}, and the scale independent NNLO coefficients $I^\two_{qi}(x)$ are given by
\begin{align}
I^{(2)}_{qq'}(x) &= C_F T_F \bigg[ -\frac{8 (1-x) (2x^2-x+2)}{3x} \big( H_{1,0} + \zeta_2 \big)
- \frac{2}{3} (8x^2+3x+3) H_{0,0} \nn
\\
&+ 4(x+1) H_{0,0,0} + \frac{4}{9} (32x^2-30x+21) H_0 + \frac{2 (1-x) (136x^2-143x+172)}{27x} \bigg] \, , \nonumber
\\
I^{(2)}_{q\bar{q}}(x) &= (C_A C_F - 2 C_F^2) \bigg[ 4(1-x) H_{1,0} + 4 (x+1) H_{-1,0} - (11x+3) H_0 + 2(3-x) \zeta_2 \nn
\\
&- 15 (1-x) - 2p_{qq}(-x) (4 H_{-2,0} - 2 H_{2,0} - 4 H_{-1,-1,0} + 2 H_{-1,0,0} - H_{0,0,0} \nn
\\
&- 2 H_{-1} \zeta_2 + \zeta _3 ) \bigg] + I^{(2)}_{qq'}(x) \, , \nonumber
\\ 
I^{(2)}_{qg}(x) &= C_A T_F \bigg[ 4 p_{qg}(-x) (2 H_{-2,0} - 2 H_{-1,-1,0} + H_{-1,0,0} - H_{-1} \zeta_2) + 8 x (x+1) H_{-1,0} \nonumber
\\
&- \frac{8 (1-x) (11x^2-x+2)}{3x} H_{1,0} + 4 p_{qg}(x) (H_{1,2}+H_{1,1,0}-H_{1,1,1}) + 8 x \zeta_3 \nonumber
\\
&- 8 (1-x) x H_{1,1} - 16 x H_{2,0} + 4 (2x+1) H_{0,0,0} + \frac{4}{9} (68x^2-30x+21) H_0 \nonumber
\\
&+ 2 x (4x-3) H_1 + \frac{8(11x^3-9x^2+3x-2)}{3x} \zeta_2 - \frac{2(298x^3-387x^2+315x-172)}{27x}\nonumber
\\
&- \frac{2}{3} (44x^2-12x+3) H_{0,0} \bigg] \nn
\\
&+ C_F T_F \bigg[ 4 p_{qg}(x) (H_{2,1}-H_{1,0,0}+H_{1,1,1}+7\zeta_3) + (-8x^2+12x+1)H_{0,0} \nn
\\
&- 2(4x^2-2x+1) H_{0,0,0} + 8 (1-x) x (H_{1,0}+H_{1,1}+H_2-\zeta_2) + (-8x^2+15x+8) H_0 \nn
\\
&- 2(4x-3) x H_1 - 72x^2 + 75x - 13 \bigg] \, , \nn
\\  
I^{(2)}_{qq}(x) &= C_A C_F \bigg[ \left( 28 \zeta_3 - \frac{808}{27} \right) \frac{1}{(1-x)_+} + 2 p_{qq}(x) (-2H_{1,2}-2 H_{2,0}-H_{0,0,0}-2H_{1,1,0}) \nn
\\ 
&+ \frac{(x^2-12x-11)}{3(1-x)} H_{0,0} - 4 (1-x) H_{1,0} - \frac{2(83x^2-36x+29)}{9(1-x)} H_0 - 2 x H_1 \nonumber
\\
&+ \frac{2(x^2-13)}{1-x} \zeta_3 - 6 (1-x) \zeta_2 + \frac{8(x+100)}{27} \bigg] \nn
\\
&+ C_F T_F N_f \bigg[ \frac{224}{27} \frac{1}{(1-x)_+} + \frac{4}{3} p_{qq}(x) H_{0,0} + \frac{20}{9} p_{qq}(x) H_0 - \frac{4}{27} (19x+37) \bigg] \nn
\\
&+ C_F^2 \bigg[ -\frac{2(2x^2-2x-3)}{1-x} H_{0,0} + 12 (1-x) H_{1,0} + \frac{2(16x^2-13x+5)}{1-x} H_0 \nn
\\
&+ 2 p_{qq}(x) (4H_{1,2}+4H_{2,0}+2H_{2,1}-2H_{1,0,0}+4H_{1,1,0}+12\zeta_3) + 2 (x+1) H_{0,0,0} \nonumber
\\
&+ 4 (1-x) H_2 + 2 x H_1 + 8 (1-x) \zeta_2 - 22 (1-x) \bigg] + I^{(2)}_{qq'}(x) \, ,
\end{align}
where $q'$ is a light quark flavor different from $q$, and we have used the shorthand notation
\begin{equation}
H_{a_1,\ldots,a_n} \equiv H(a_1,\ldots,a_n;x) \, ,
\end{equation}
with $H$ being HPLs. The $p_{qi}(x)$ functions are related to the DGLAP splitting kernels and are collected in the Appendix. We note that with the exponential regulator, the scale-independent part of the TMDPDF does not involve $\delta(1-x)$ terms, and the coefficients of $1/(1-x)_+$ is determined by the rapidity anomalous dimension $2\gamma_1^R$ given in Eq.~\eqref{eq:gammaR}. We have compared our results to those in the literature~\cite{Gehrmann:2012ze,Gehrmann:2014yya,Echevarria:2016scs} and found full agreement. We have also obtained the bare NNLO TMDPDFs through to $\cO(\e^2)$, which is required for a future N$^3$LO calculation. Their expressions are quite lengthy and we choose to put them in an electronic file attached with the arXiv submission of this paper.

\section{Quark TMDFF with the exponential regulator}
\label{sec:tmdff}

We now turn to the quark TMDFF. Technically, it is very similar to the TMDPDF. The squared amplitudes are related via a crossing symmetry. The only subtlety is that one may perform the calculations in the hadron frame or in the parton frame. The two results should be related according to Eqs.~\eqref{eq:ff1_parton_hadron} and \eqref{eq:ff2_parton_hadron}. We have explicitly performed the two calculations and confirmed those relations.

In the hadron frame, one has, at a given order in perturbation theory
\begin{multline}
\mathcal{D}^{\text{bare}}_{i/q}(z,b_\perp,\nu) = \lim_{\tau \to 0} \frac{1}{z} \int d^dk \, e^{-b_0 \tau k^0 + i \bp \cdot \vec{k}_{T,\text{hf}}}  \, \delta(k_- - (1/z-1)p_-)
\\ 
\times \prod \int \frac{d^dl_i}{(2\pi)^d} \prod \int \frac{d^dk_i}{(2\pi)^d}  \, (2\pi) \delta_{+}(k_i^2) \, \delta^{(d)}\bigg(k-\sum_{i=1}^n k_i \bigg) \, \mathcal{M}_{iq}(p,l_i,k_i) \bigg|_{\tau=1/\nu} \, ,
\end{multline}
where $\mathcal{M}_{iq}$ is the squared amplitude for the $q \to i$ splitting, $p$ is the momentum of the observed hadron in the $\bar{n}$ direction, $l_i$ are loop momenta and $k_i$ are momenta of real emissions, $\vec{k}_{T,\text{hf}}$ denotes the total transverse momentum of real emissions in the hadron frame.
In the parton frame, one has instead
\begin{multline}
\mathcal{F}^{\text{bare}}_{i/q}(z,b_\perp/z,\nu) = \lim_{\tau \to 0} \frac{1}{z} \int d^dk \, e^{-b_0 \tau k^0 + i \bp \cdot \vec{k}_{T,\text{pf}} / z}  \, \delta(k_- - (1/z-1)p_-)
\\ 
\times \prod \int \frac{d^dl_i}{(2\pi)^d} \prod \int \frac{d^dk_i}{(2\pi)^d}  \, (2\pi) \delta_{+}(k_i^2) \, \delta^{(d)}\bigg(k-\sum_{i=1}^n k_i \bigg) \, \mathcal{M}_{iq}(p,l_i,k_i) \bigg|_{\tau=1/\nu} \, ,
\end{multline}
where $\vec{k}_{T,\text{pf}}$ denotes the total transverse momentum of real emissions in the parton frame.
The only differences with respect to the hadron frame formula are the additional factor of $1/z$ in the Fourier transform, and the different definition of $\kp$. They are related by $\vec{k}_{T,\text{pf}} = z \vec{k}_{T,\text{hf}}$.

\subsection{Quark TMDFF at NLO and NNLO}

\begin{figure}[t!]
\centering
\includegraphics[width=0.5\linewidth]{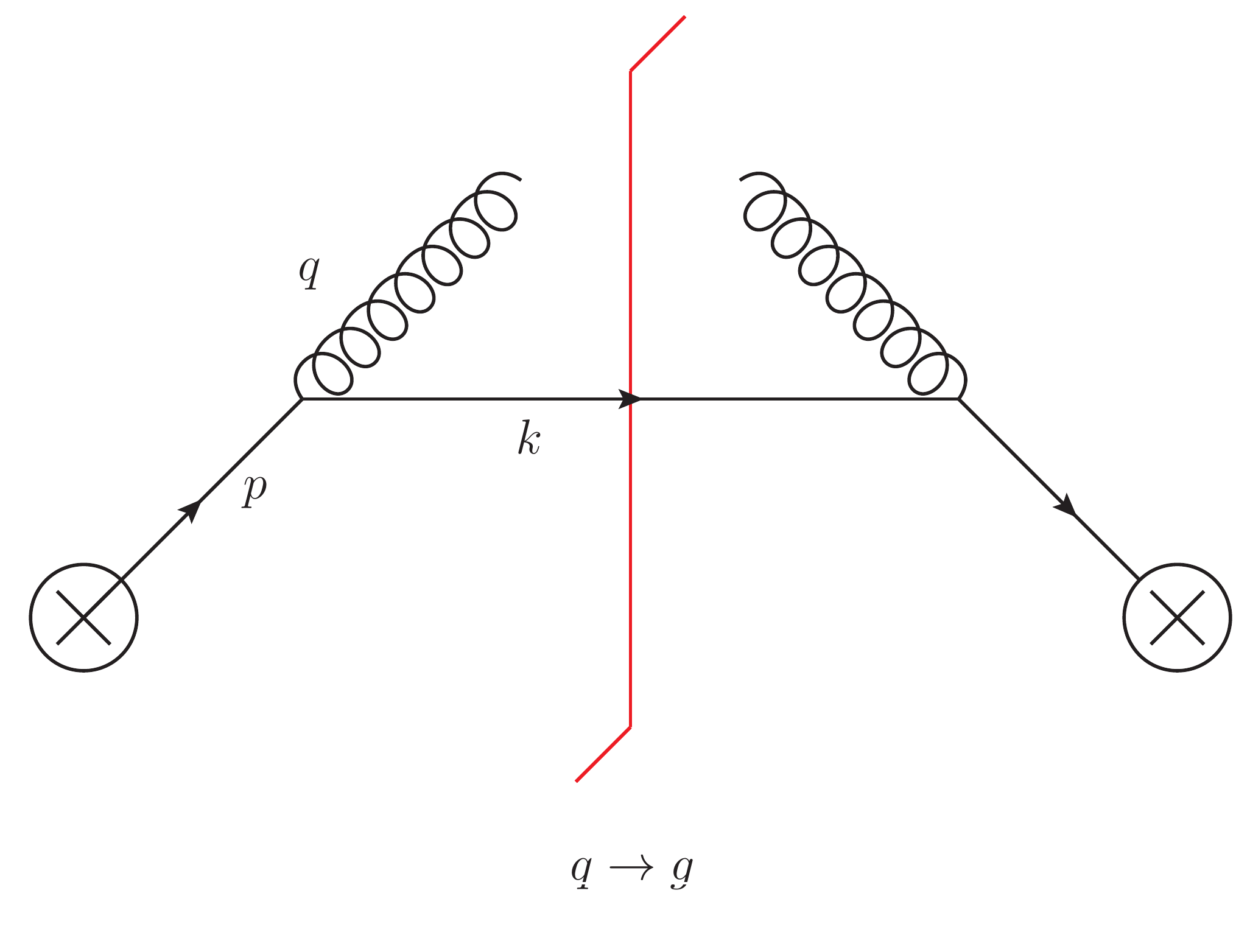}
\caption{\label{fig:NLOTMDFFgq}Cut diagram for the bare $q\rightarrow g$ TMDFF at NLO.}
\end{figure}

We now present some details about the calculation of the $q \to g$ fragmentation function at NLO in the hadron frame. In the light-cone gauge, there is only one cut diagram contributing, as shown in Fig.~\ref{fig:NLOTMDFFgq}. The squared amplitude can be straightforwardly obtained from the diagram, or can be related to the $g \to q$ splitting amplitude in Eq.~\eqref{eq:splitting} via a crossing symmetry. We have
\begin{equation}
\mathcal{M}_{gq}(p,k) = \frac{g_s^2\mu^{2\e}k_-}{z^2k_T^2} \, p_{gq}^\zero(z,\e) \, ,
\end{equation}
with
\begin{equation}
p_{gq}^\zero(z,\e) = \frac{1 - \epsilon}{z} \, p_{qg}^\zero(1/z,\e) \Big|_{T_F \rightarrow C_F} = 2 C_F \left[ \frac{1 + (1-z)^2}{z} - \e z \right] .
\label{eq:splittinggq}
\end{equation}
The above equation is the manifestation of the Gribov-Lipatov relation.

For the $q \to g$ fragmentation at NLO, there are no rapidity divergences, and therefore we do not need to introduce the exponential regulator. The bare TMDFF in the hadron frame then reads
\begin{multline}
\alspi \mathcal{D}_{g/q}^{\one,\text{bare}}(z,b_\perp,\nu) = \int \frac{\df^dk}{(2\pi)^d} \, (2\pi) \delta_+(k^2) \, \delta(k_- - (1/z-1) p_{-}) \, e^{i \bp \cdot \kp} \,  \frac{g_s^2 \mu^{2\e} k_-}{z^2\kpsq} \, p_{gq}^\zero(z,\e) \, .
\end{multline}
The above integral is similar to the ones appearing in the calculation of TMDPDFs. The result is
\begin{align}
\mathcal{D}_{g/q}^{\one,\text{bare}}(z,b_\perp,\nu) &= \frac{2C_F}{z^2} \left[ -\frac{1+(1-z)^2}{z} \left( \frac{1}{\e} + \Lp \right) + z \right] + \mathcal{O}(\e) \nn
\\
&\equiv \frac{2C_F}{z^2} \left[ -p_{gq}(z) \left( \frac{1}{\e} + \Lp \right) + z \right] + \mathcal{O}(\e) \, .
\end{align}
We now need to proceed with the matching procedure \eqref{eq:TMDFFOPE}, where one should pay attention to the prefactor $z^{2-2\e}$, which will produce logarithms of $z$ when expanding in $\e$. We have
\begin{equation}
\mathcal{D}_{g/q}^{\one,\text{bare}}(z,b_\perp,\nu) \, z^{2-2\e} = 2C_F \left[ -p_{gq}(z) \left( \frac{1}{\e} + \Lp - 2\ln z \right) + z \right] + \mathcal{O}(\e) \, .
\end{equation}
Performing matching and renormalization as in Eq.~\eqref{eq:TMDFFOPE}, we then obtain
\begin{equation}
\cC^\one_{gq}(z,b_\perp/z,L_Q) = 2C_F \left[ -p_{gq}(z) \Lp + 2p_{gq}(z) \ln z + z \right] .
\end{equation}
Note that the scale-dependent part agrees with Eq.~\eqref{eq:RGexpC}.

We perform the calculation for the $q \to q$ fragmentation in a similar manner, where we need to use the exponential regulator for the rapidity divergences. The scale-independent coefficients at NLO are then given by 
\begin{align}
C^{(1)}_{qq}(z) &= 2 C_F \left( 2 p_{qq}(z) H_0 + 1-z  \right), \nonumber \\
C^{(1)}_{gq}(z) &= 2 C_F \left( 2 p_{gq}(z) H_0 + z \right),
\end{align}
where we use the shorthand notation
\begin{equation}
H_{a_1,\ldots,a_n} \equiv H(a_1,\ldots,a_n;z) \, .
\end{equation}

The NNLO calculations proceed in an analogous way, and we do not repeat the details here. The results are
\begin{align}
C_{q'q}^{(2)}(z) &= C_F T_F \bigg[ \frac{8 (1-z) (2z^2-z+2)}{3z} (H_{1,0}+\zeta_2) - \frac{2(24z^3+33z^2+33z-32)}{3z} H_{0,0} \nonumber
\\
&- \frac{4(32z^3+51z^2+174z-12)}{9z} H_0 - \frac{2 (1-z) (436z^2+859z+148)}{27z} \nonumber \\
&+44 (z+1) H_{0,0,0} \bigg] \, , \nn
\\
C_{\bar{q}q}^{(2)}(z) &= C_{q'q}^{(2)}(z) + (C_A C_F-2 C_F^2) \bigg[ -4 (1-z) H_{1,0} + 4 (z+1) H_{-1,0} - 8 (z+2) H_{0,0} \nonumber
\\ 
&- 2 p_{qq}(-z) ( -4 H_{-2,0} + 2H_{2,0} - 4 H_{-1,-1,0} - 6 H_{-1,0,0} + 9 H_{0,0,0} - 2 H_{-1} \zeta_2 + 3 \zeta_3 ) \nonumber
\\
&+ (5z-19) H_0 + 2 (3z-1) \zeta_2 - 15 (1-z) \bigg] \, , \nonumber
\\
C_{gq}^{(2)}(z) &= C_A C_F \bigg[ -4p_{gq}(-z) ( 2 H_{-2,0} + 2 H_{-1,-1,0} + 3 H_{-1,0,0} + H_{-1} \zeta_2 ) \nn
\\
&+ 4 p_{gq}(z) (-4 H_{1,2} - 3 H_{2,1} - 11 H_{1,0,0} - 4 H_{1,1,0} - H_{1,1,1} + 4 H_0 \zeta_2 + 3 H_1 \zeta_2 - 2 H_3) \nonumber
\\
&- \frac{8 (5z^2-8z+10)}{z} H_{2,0} - \frac{4 (31z^2+22z+40)}{z} H_{0,0,0} + 4 z ( H_{-1,0} - H_{1,1} + H_2) \nn
\\
&+\frac{2 (24z^3-9z^2+96z-212)}{3z}  H_{0,0} + 2 H_1 + \frac{4 (4z^3-15z^2+24z-22)}{3z} H_{1,0} \nonumber
\\
&+ \frac{2 (88z^3+147z^2+735z+54)}{9z} H_0 - \frac{2 (340z^3+693z^2+558z-1564)}{27z}\nonumber
\\
&- \frac{8 (1-z) (2z^2-z+11)}{3z} \zeta_2 - \frac{4 (11z^2-16z+22)}{z} \zeta_3 \bigg] \nn
\\
&+C_F^2 \bigg[ -8 z H_{1,0} + 4 z H_{1,1} + (-z-8) H_{0,0} - 22 (z-2) H_{0,0,0} - 8 z H_2 - 2 H_1 \nonumber
\\
&+ 4 p_{gq}(z) (3 H_{1,2} - 4 H_{2,0} + 2 H_{2,1} + 4 H_{1,0,0} + 3 H_{1,1,0} + H_{1,1,1} - 8 H_0 \zeta_2 - 3 H_1 \zeta_2 \nn
\\
&\hspace{4em} - 4 H_3 ) - \frac{(z+3) (13z-16)}{z} H_0 - 16 z \zeta_2 + 33 z - 38 \bigg] \, , \nonumber
\\
C_{qq}^{(2)}(z) &= C_{q'q}^{(2)}(z) + C_A C_F \bigg[ \left( 28 \zeta_3 - \frac{808}{27} \right) \frac{1}{(1-z)}_+ - \frac{(23z^2+36z-37)}{3(1-z)} H_{0,0} \nn
\\
&+ 4 (1-z) H_{1,0} + 2p_{qq}(z) (2 H_{1,2} + 9 H_{0,0,0} + 4 H_{1,0,0} + 2 H_{1,1,0} - 6 H_0 \zeta_2 + 2 H_3 ) \nonumber
\\
&+ \frac{2 (72z^2-95z+93)}{3(1-z)} H_0 + 2 H_1 + \frac{2 (3z^2-11)}{1-z} \zeta_3 + 2 (1-z) \zeta_2 + \frac{8 (z+100)}{27} \bigg] \nonumber
\\ 
&+ C_F T_F N_f \bigg[ \frac{224}{27} \frac{1}{(1-z)}_+ + \frac{4}{3} p_{qq}(z) H_{0,0} -\frac{4 (9z^2-8z+9)}{3(1-z)} H_0 - \frac{4}{27} (19z+37) \bigg] \nonumber
\\
&+ C_F^2 \bigg[ \frac{2 (3z^2+34z-22)}{1-z} H_{0,0} - \frac{2 (51z^2+29)}{1-z} H_{0,0,0} - 24 (1-z) \zeta_2 + 10 (1-z) \nonumber
\\
&+ 2p_{qq}(z) (-4 H_{1,2} - 26 H_{2,0} - 2 H_{2,1} - 18 H_{1,0,0} - 4 H_{1,1,0} - 2 H_0 \zeta_2 - 14 H_3 - 22 \zeta_3 ) \nonumber
\\
&- 28 (1-z) H_{1,0} - \frac{2 (27z^2-42z+23)}{1-z} H_0 - 4 (1-z) H_2 - 2 H_1 \bigg] \, .
\end{align}
Again, we find that the scale independent parts do not contain $\delta(1-z)$ terms, and the $1/(1-z)_+$ terms are determined by the rapidity anomalous dimension. We can convert our results to the convention of Ref.~\cite{Echevarria:2015usa} and compare with the results in that work. We find that the results agree for the splitting processes $q \to q'$, $q \to \bar{q}$ and $q \to g$. However, for $q \to q$, there is a small difference concerning a term $C_A C_F \, \pi^4 \, \delta(1-z)$. In our framework, this term comes from the TMD soft function, which is universal for the TMDPDF and TMDFF. To address this discrepancy, we have performed several independent checks. The strongest check of our calculation comes from the calculation of the two-loop jet function of EEC in the back-to-back limit, which we shall explain in the next subsection.

\subsection{Jet function for the EEC in the back-to-back limit}
\label{sec:jet-function-eec}

The EEC measures the energy correlation of two detectors in $e^+e^-$ annihilation at an angle $\chi$.
The TMDFFs obtained in the last subsection can be used to calculate the jet function for the EEC in the back-to-back limit $\chi \to \pi$. It has been known for a long time that resummation of large logarithms for the EEC in this limit is closely related to $q_T$ resummation in the Drell-Yan process~\cite{Collins:1981uk,Dokshitzer:1999sh,deFlorian:2004mp}. Recently, an all-order factorization formula in terms of operator matrix elements for the EEC in the back-to-back limit has been presented~\cite{Moult:2018jzp}. The factorization formula at leading power reads
\begin{align}
\label{eq:EEC}
\frac{1}{\sigma_0} \frac{d\sigma}{dz}= \int d\bpsq \, J_0(b_T Q \sqrt{1-z}) \, H(Q,\mu) \,
J^q (b_\perp,\mu,\nu) \, J^{\bar{q}}(b_\perp,\mu,\nu) \, S(b_\perp,\mu,\nu) \,,
\end{align}
where $z = (1 - \cos\chi)/2$. In the back-to-back limit one has $z \to 1$. The hard function and soft function in Eq.~\eqref{eq:EEC} is well known and can be found to two loops in Ref.~\cite{Moult:2018jzp}. The only missing ingredient for the resummation at N$^3$LL accuracy is the two-loop jet function. In QCD, due to charge conjugation invariance, we have $J^q(b_\perp,\mu,\nu) = J^{\bar{q}}(b_\perp,\mu,\nu)$. The quark jet function can be obtained from the second Mellin moments of the matching coefficients of quark TMDFFs~\cite{Moult:2018jzp}
\begin{align}
J^q(b_\perp,\mu,\nu) = \sum_i \int_0^1 dx\, x \, \cC_{iq}(x, b_\perp/x, \mu, \nu) \,.
\end{align}
We expand the jet function in terms of $\alpha_s$ as
\begin{align}
J^q (b_\perp,\mu,\nu) = \sum_{n=0} \left( \alsmupi \right)^n J^q_n (b_\perp,\mu,\nu) \,.
\end{align}
Using the two-loop TMDFFs computed in this paper, the expansion coefficients are then given by
\begin{align}
J_0^q &= \int_0^1 dx\, x \, \cC_{qq}^{(0)}(x, b_\perp/x, \mu, \nu) = 1 \,,
\nn\\
J_1^q &= \int_0^1 dx\, x \left( \cC_{qq}^{(1)} + \cC_{gq}^{(1)} \right)=  C_F \left( -2 \Lp L_Q + 3 \Lp - 8 \zeta_2 + 4 \right) ,
\nn \\
J_2^q &= \int_0^1 dx\, x \, \left( \cC_{qq}^{(2)} + \cC_{gq}^{(2)} + \cC_{\bar{q}q}^{(2)} + 2 (N_f - 1) \cC_{q'q}^{(2)} \right)
\nn\\
&= C_A C_F \bigg[ \Lp \left( \left( 4 \zeta_2 - \frac{134}{9} \right) L_Q - \frac{44\zeta_2}{3} - 12\zeta_3 + \frac{35}{2} \right) + \Lp^2 \left( \frac{11}{2} - \frac{11}{3} L_Q \right) \nn \\
&\hspace{4em} + \left( 14\zeta_3 - \frac{404}{27} \right) L_Q \bigg]
\nn\\
&+ C_F T_F N_f \bigg[ \Lp \left( \frac{40}{9} L_Q + \frac{16\zeta_2}{3} - 6 \right) + \Lp^2 \left( \frac{4}{3} L_Q - 2 \right) + \frac{112}{27} L_Q \bigg]
\nn\\
&+C_F^2 \bigg[ \Lp \left( (16\zeta_2-8) L_Q - 36\zeta_2 + 24\zeta_3 + \frac{27}{2} \right) + \Lp^2 \left( 2L_Q^2 - 6L_Q + \frac{9}{2} \right) \bigg] + c_2^J \,.
\label{eq:twoloopjet}
\end{align}
The $\mu$ and $\nu$ dependence of the jet function are in full agreement with the RGE and rapidity evolution equation~\cite{Moult:2018jzp}. The new result from this paper is the two-loop constant term
\begin{align}
c_2^J &= C_A C_F \left( -\frac{178 \zeta_2}{3} + \frac{74 \zeta_3}{3} - 5\zeta_4 + \frac{1549}{72} \right) + C_F T_F N_f \left( \frac{56 \zeta_2}{3} + \frac{8\zeta_3}{3} - \frac{149}{18} \right) \nn
\\ 
&+ C_F^2 \left( -28 \zeta_2 - 74 \zeta_3 + 140 \zeta_4 + \frac{139}{24} \right) \,,
\end{align}
which represents the last missing ingredient for N$^3$LL resummation of EEC in the back-to-back limit.

Using the two-loop jet function together with the two-loop hard and soft function, we obtain the full leading power prediction for the EEC in the back-to-back limit through two loops from the factorization formula in Eq.~\eqref{eq:EEC}, including the $\delta(1-z)$ terms,
\begin{align}
  \frac{1}{\sigma_0} \frac{d\sigma^{(0)}}{dz}\bigg|_{z\to 1} &= \frac{1}{2}\delta(1-z)  \,,
\nn\\
  \frac{1}{\sigma_0} \frac{d\sigma^{(1)}}{dz}\bigg|_{z\to 1} &= 
C_F \left[ (-2 \zeta_2 - 4) \delta(1-z) - 3 D_0(z) - 2 D_1(z) \right] ,
\nn\\
  \frac{1}{\sigma_0} \frac{d\sigma^{(2)}}{dz}\bigg|_{z\to 1} &=
C_A C_F \left[ D_0(z) \left(22 \zeta_2+12
   \zeta_3-\frac{35}{2}\right)+D_1(z) \left(4
   \zeta_2-\frac{35}{9}\right)+\frac{22}{3} D_2(z) \right]
\nn\\
&+ C_F N_f
   \left[ D_0(z) (3-4 \zeta_2)+\frac{2}{9}D_1(z)-\frac{4}{3}D_2(z) \right]
\nn\\
&+ C_F^2 \left[ D_0(z) \left(24 \zeta_2-8
   \zeta_3+\frac{45}{2}\right)+D_1(z) (8 \zeta_2+34)+18 D_2(z)+4
   D_3(z)\right]
\nn\\
&\ + c_2^{z=1}\delta(1-z) \,,
\label{eq:endpoint}
\end{align}
where we have set $\mu = Q$ and $T_F = 1/2$ for simplicity, and
\begin{align}
D_n(z) \equiv \left[\frac{\ln^n (1-z)}{1-z} \right]_+   \,.
\end{align}
The two-loop $\delta(1-z)$ term is
\begin{align}
  \label{eq:deltaomz}
  c_2^{z=1} &= 
C_A C_F \left(-\frac{104 \zeta_2}{9}+\frac{182 \zeta_3}{3}-8
   \zeta_4-\frac{382}{9}\right)+C_F N_f \left(\frac{8 \zeta_2}{9}+\frac{4
   \zeta_3}{3}+\frac{58}{9}\right)
\nn\\
&+C_F^2 \left(49 \zeta_2-80 \zeta_3+48
   \zeta_4+\frac{41}{3}\right) \,.
\end{align}
The two-loop plus distribution terms $D_n(z)$ are in full agreement with the analytical NLO calculation in Ref.~\cite{Dixon:2018qgp}, while the two-loop $\delta(1-z)$ term is new. Eq.~\eqref{eq:deltaomz} has already been used in a previous publication to extract the $\delta(z)$ term of EEC using the energy conservation sum rule~\cite{Dixon:2019uzg}. Two independent checks are made for Eq.~\eqref{eq:deltaomz}. Firstly, EEC in the back-to-back limit obey the leading transcendental principle~\cite{Korchemsky:2019nzm,Henn:2019gkr}, which states that the maximal transcendental part of the QCD results are identical to the same quantity in ${\cal N}=4$ supersymmetric Yang-Mills (SYM) theory, up to trivial overall color factor~\cite{Kotikov:2002ab}. In Eq.~\eqref{eq:deltaomz}, the leading transcendental term is the $\zeta_4$ terms. To compare with the same quantity in ${\cal N}=4$ SYM theory, we replace $C_F \to C_A$ in Eq.~\eqref{eq:deltaomz} and found the leading transcendental piece to be $40 C_A^2 \zeta_4$, which is in full agreement with an independent calculation in~\cite{Korchemsky:2019nzm}. Secondly, Besides the energy conservation sum rule, EEC in massless perturbation theory also obey a sum rule due to momentum conservation, which reads~\cite{Kologlu:2019mfz,Korchemsky:2019nzm}
\begin{align}
  \label{eq:momsumrule}
  \frac{1}{\sigma_{\rm tot}} \int_0^1 dz \, z \frac{d \sigma}{dz} = \frac{1}{2} \,,
\end{align}
where $\sigma_{\rm tot}$ is the total hadronic cross section for $e^+e^-$ including higher order QCD corrections. Using the analytical NLO formula of EEC for $0<z<1$ from \cite{Dixon:2018qgp}, and the end point contribution in Eq.~\eqref{eq:endpoint}, we explicit verify the sum rule in Eq.~\eqref{eq:momsumrule}.

The end point contributions in Eq.~\eqref{eq:endpoint} are directly computed using the two-loop jet function in Eq.~\eqref{eq:twoloopjet}, which by itself reduces to moment of the TMDFFs. Therefore, the checks made for the end point contributions apply also to the TMDFFs computed in this paper, in particular to its $\delta(1-z)$ terms.  

\section{Conclusion}
\label{sec:conclusion}

In this work, we have revisited the calculation of perturbative quark TMDPDFs and TMDFFs at NNLO using a new regulator for rapidity divergences. We use the SIDIS process to set-up our calculation, while our results are universal and can be used for other processes as well. We show that the exponential regulator provides a consistent framework to carry out the calculation of the TMD soft functions, TMDPDFs and TMDFFs.

Compared to existing regulators in the literature, the exponential regulator has a couple of advantages. Firstly, the regulator can be implemented at the level of operator definitions for the TMD functions, where it manifests itself as a small shift of the space-time coordinates. Secondly, the exponential regulator is applied to the total momentum of the extra emissions in the final state. Except for this last integration, the regulator does not change the structure of (cut)-propagators in the amplitudes. As a result, we can apply many modern techniques for loop integrals such as IBP identities and differential equations. This allows us to obtain the bare NNLO TMDPDFs and TMDFFs up to $\Ord(\e^2)$ in dimensional regularization, and can also be extended to a future N$^3$LO calculation. Finally, the regulator can already be expanded in terms of delta-functions and plus-distributions at the integrand level, which makes the final round of integration easy to carry out.

Our results for the quark TMDPDFs up to $\Ord(\e^0)$ agree with the results in the literature, while our results for the quark TMDFFs have a small discrepancy with another calculation. To further check our results, we use the TMDFFs to calculate the NNLO jet function appearing in the factorization formula of EEC in the back-to-back-limit. This also serves as a new result of our paper, and is the last missing ingredient for an N$^3$LL resummation. We have checked that our NNLO jet function produces the correct leading singular terms for the EEC in the back-to-back limit. This is a strong validation of our results for the TMDFFs.

Given the benefits provided by the exponential regulator, the calculation for the gluon TMDPDFs and TMDFFs (which was more difficult than the quark case using other regulators) can also be greatly simplified. We also believe that our method can be extended to the N$^3$LO level. We leave these considerations to future publications.

\acknowledgments

This work was supported in part by the National Natural Science Foundation of China (11135006, 11275168, 11422544, 11375151, 11535002, 11575004, 11635001), and the Zhejiang University
Fundamental Research Funds for the Central Universities (2017QNA3007).

\appendix

\section{Anomalous dimensions, splitting functions and the TMD soft function}
\label{sec:appendix}

In this Appendix, we list some necessary ingredients which enter our calculation.

\subsection{Anomalous dimensions}

For all the anomalous dimensions entering the RGEs of various TMD functions, we define the perturbative expansion according to
\begin{equation}
\gamma(\als) = \sum_{n=0}^\infty \left( \alspi \right)^{n+1} \, \gamma_n \, ,
\end{equation}
where the coefficients up to $\Ord(\alpha_s^2)$ are
\begin{align}
\Gcusp_{0} &= 4 C_F \nn
\\
\Gcusp_{1} &= C_A C_F \left(\frac{268}{9}-8 
                 \zeta_2\right)-\frac{80 C_F T_F N_f}{9} \nn
\\
\gamma^B_0 &= 3C_F \, , \nn
\\
\gamma^B_1 &= C_F \left[ C_F \left( \frac{3}{2} - 2\pi^2 + 24\zeta_3 \right) + C_A \left( \frac{17}{6} + \frac{22\pi^2}{9} - 12\zeta_3 \right) + T_FN_f \left( -\frac{2}{3} - \frac{8\pi^2}{9} \right) \right] , \nn
\\
\gamma^H_0 &= -3C_F \, , \nn
\\
\gamma^H_1 &= C_F \left[ C_F \left( -\frac{3}{2} + 2\pi^2 - 24\zeta_3 \right) + C_A \left( -\frac{961}{54} - \frac{11\pi^2}{6} + 26\zeta_3 \right) + T_FN_f \left( \frac{130}{27} + \frac{2\pi^2}{3} \right) \right] , \nn
\\
\gamma^S_0 &= 0 \, , \nn
\\
\gamma^S_1 &= C_F \left[ C_A \left( -\frac{404}{27} + \frac{11\pi^2}{18} + 14\zeta_3 \right) + T_FN_f \left( \frac{112}{27} - \frac{2\pi^2}{9} \right) \right] .
\end{align}
The cusp anomalous dimension $\Gamma^{\text{cusp}}$ can be found in \cite{Korchemsky:1987wg,Korchemskaya:1992je}. The hard and soft anomalous dimensions $\gamma^H$ and $\gamma^S$ can be extracted from the two-loop quark form factor~\cite{Gehrmann:2005pd, Moch:2005id}, and can also be found in, e.g., Refs.~\cite{Becher:2009qa,Li:2014afw}. Finally, the beam anomalous dimension $\gamma^B$ is related to $\gamma^S$ and $\gamma^H$ through $\gamma^B = \gamma^S - \gamma^H$. And the renormalization factor for the quark TMDPDFs and TMDFFs up to $\mathcal{O}(\alpha_s^2)$ reads
\begin{align}
	\label{eq:R_factor}
	Z_q^B(b_\perp,\mu,\nu) &= 1 +\frac{\alpha_s}{4\pi}\frac{\Gcusp_0 L_Q-2\gamma^B_0}{-2\e}\nn \\
	& +\left(\frac{\alpha_s}{4\pi}\right)^2\left[\frac{(\Gcusp_0 L_Q-2\gamma^B_0)^2+2\beta_0(\Gcusp_0 L_Q-2\gamma^B_0)}{8\e^2}+\frac{\Gcusp_1 L_Q-2\gamma^B_1}{-4\e}\right].
\end{align}

The QCD beta function is defined by
\begin{equation}
\frac{d\als}{d\ln\mu} = \beta(\als) = -2\als \sum_{n=0}^\infty \left( \alspi \right)^{n+1} \, \beta_n \, ,
\end{equation}
with \cite{Gross:1973id,Politzer:1973fx,Caswell:1974gg,Jones:1974mm,Egorian:1978zx}
\begin{align}
\beta_0 &= \frac{11}{3} C_A - \frac{4}{3} T_F N_f \, , \nn
\\
\beta_1 &= \frac{34}{3} C_A^2 - \frac{20}{3} C_A T_F N_f - 4 C_F T_F N_f \, ,
\end{align}
A formula particularly useful for us is
\begin{equation}
\als(b_0/b_T) = \frac{\alsmu}{t} \left[ 1 - \alsmupi \frac{\beta_1}{\beta_0} \frac{\ln t}{t} \right] + \Ord(\als^3) \, ,
\end{equation}
where
\begin{equation}
t = 1 - \alsmupi \beta_0 \Lp \, .
\end{equation}

\subsection{Space-like splitting functions}

The first order space-like splitting functions can be written as \cite{Altarelli:1977zs}
\begin{align}
P^{\zero}_{qq}(z) &= 2 C_F \left[ p_{qq}(z) +\frac{3}{2} \delta (1-z) \right] , \nonumber  \\
P^{\zero}_{gq}(z) &= 2 C_F \, p_{gq}(z) \, , \nonumber \\
P^{\zero}_{gg}(z) &= 4 C_A \, p_{gg}(z) + \delta(1-z) \left( \frac{11}{3}  C_A - \frac{4}{3} T_F N_f \right) , \nonumber \\
P^{\zero}_{qg}(z) &=  2 T_F \, p_{qg}(z) \, ,
\end{align}
where
\begin{align}
p_{qq}(z) &= \frac{1+z^2}{(1-z)_+} \, , \nonumber \\
p_{gq}(z) &= \frac{1+(1-z)^2}{z} \, , \nonumber \\
p_{gg}(z) &= \frac{z}{(1-z)_+} + \frac{1-z}{z} + z(1-z) \, , \nonumber \\
p_{qg}(z) &= z^2+ (1-z)^2 \, .
\end{align}
The second order space-like splitting functions are \cite{Furmanski:1980cm,Curci:1980uw}
\begin{align}
P^{\one}_{qq'}(z) &= C_F T_F \bigg[ -8 (z+1) H_{0,0} + \frac{4}{3} (8z^2+15z+3) H_0  + \frac{8 (1-z)
(28z^2+z+10)}{9z} \bigg] \, , \nonumber
\\
P^{\one}_{q\bar{q}}(z) &= P^{\one}_{qq'}(z) + (C_A C_F - 2 C_F^2) \bigg[ 4 p_{qq}(-z) (2 H_{-1,0} - H_{0,0} + \zeta_2) - 4 (z+1) H_0 - 8 (1-z) \bigg] \, , \nonumber
\\  
P^{\one}_{qg}(z) &= C_A T_F \bigg[ -8 p_{qg}(-z) H_{-1,0} - 8 p_{qg}(z) H_{1,1} - 8 (2z+1) H_{0,0} + 16 (1-z) z H_1 \nonumber \\ 
&+ \frac{4}{3} (44z^2+24z+3) H_0 - \frac{4(218z^3-225z^2+18z-20)}{9z} - 16 z \zeta_2 \bigg] \nonumber \\
&+ C_F T_F \bigg[ 8 p_{qg}(z) (H_{1,0}+H_{1,1}+H_2-\zeta_2) + 4 (4z^2-2z+1) H_{0,0} \nonumber \\
&+ 2 (8z^2-4z+3) H_0 - 16 (1-z) z H_1 + 2 (20z^2-29z+14) \bigg] \, , \nonumber
\\
P^{\one}_{qq}(z) &= P^{\one}_{qq'}(z) + C_A C_F \bigg[ \left(\frac{268}{9}-8\zeta_2\right) \frac{1}{(1-z)_+} + 4 p_{qq}(z) H_{0,0} + \frac{2(5z^2+17)}{3(1-z)} H_0 \nn \\
&+ 4(z+1) \zeta_2 + \left( \frac{44\zeta_2}{3} - 12 \zeta_3 + \frac{17}{6} \right) \delta(1-z) - \frac{2}{9} (187z-53) \bigg] \nn
\\
&+ C_F T_F N_f \bigg[ -\frac{80}{9} \frac{1}{(1-z)}_+  - \frac{8}{3} p_{qq}(z) H_0 + \left( -\frac{16\zeta_2}{3} - \frac{2}{3} \right) \delta(1-z) + \frac{8}{9} (11z-1) \bigg] \nn \\
&+ C_F^2 \bigg[ 8 p_{qq}(z) (H_{1,0}+H_2) - 4 (z+1) H_{0,0} + \frac{4(2z^2-2z-3)}{1-z} H_0 \nn \\
&+ \left( -12 \zeta_2 + 24 \zeta_3 + \frac{3}{2} \right) \delta(1-z) - 20(1-z) \bigg] \, .
\end{align}

\subsection{Time-like splitting function}

The first order time-like splitting functions are exactly the same as the space-like ones, while the second order time-like splitting functions are given by \cite{Furmanski:1980cm, Curci:1980uw}
\begin{align}
P^{T \one}_{q'q}(z) &= C_F T_F \bigg[ 8 (z+1) H_{0,0} - \frac{4}{3} (8z^2+27z+15) H_0 - \frac{16 (1-z)
(14z^2+23z+5)}{9z} \bigg] \, , \nonumber
\\
P^{T \one}_{\bar{q}q}(z) &=  P^{T \one}_{q'q}(z) + (C_A C_F - 2 C_F^2) \bigg[ 4 p_{qq}(-z) (2 H_{-1,0} - H_{0,0} + \zeta_2) - 4 (z+1) H_0 - 8 (1-z) \bigg] \, , \nonumber
\\
P^{T \one}_{gq}(z) &= C_A C_F \bigg[ -8p_{gq}(-z) H_{-1,0} - \frac{8(3z^2+2z+4)}{z} H_{0,0} + 8 p_{gq}(z) (-3H_{1,0}-H_{1,1}+H_2) \nonumber \\
&+ \frac{4(8z^3+27z^2+24z-18)}{3z} H_0 + 8z H_1 - \frac{4(44z^3+9z^2-45z-17)}{9z} + 16 \zeta_2 \bigg] \nonumber \\
&+ C_F^2 \bigg[ 8p_{gq}(z) (2H_{1,0}+H_{1,1}-2H_2) - 4 (z-2) H_{0,0} + 2 (z-16) H_0 - 8 z H_1 \nonumber \\
&+ 2(9z-1) \bigg] \, , \nonumber   
\\
P^{T \one}_{qq}(z) &= P^{T \one}_{q'q}(z)  + C_A C_F \bigg[ \left( \frac{268}{9} - 8 \zeta_2 \right) \frac{1}{(1-z)_+} + 4 p_{qq}(z) H_{0,0} + \frac{2 (5z^2+17)}{3(1-z)}  H_0 \nonumber \\ 
&+ \left( \frac{44}{3} \zeta_2 - 12 \zeta_3 + \frac{17}{6} \right) \delta(1-z) + 4(z+1)\zeta_2 - \frac{2}{9} (187z-53) \bigg] \nonumber \\
&+ C_F T_F N_f \bigg[ -\frac{80}{9} \frac{1}{(1-z)_+} - \frac{8}{3} p_{qq}(z) H_0 + \left( -\frac{16\zeta_2}{3} - \frac{2}{3} \right) \delta(1-z) + \frac{8}{9} (11z-1) \bigg] \nonumber \\ 
&+ C_F^2 \bigg[ -\frac{4 (5z^2+3)}{1-z} H_{0,0} + 8p_{qq}(z) (-H_{1,0}-H_2) + \frac{4(3z^2+2z-2)}{1-z} H_0 \nonumber \\
&+ \left( -12\zeta_2 + 24\zeta_3 + \frac{3}{2} \right) \delta(1-z) - 20 (1-z) \bigg] \, ,  
\end{align}
where we use the same convention as in Ref.~\cite{Ritzmann:2014mka}.

\subsection{TMD soft function}

The exponentially regularized TMD soft function is given by \cite{Li:2016ctv}
\begin{multline}
S_{q\bar{q}}(b_\perp,\mu,\nu) =  \exp \Bigg\{ \alspi \bigg[ \frac{\Gcusp_0}{2} \Lp^2 - \Lp \big( \Gcusp_0 L_R +2\gamma_0^S \big) + 2 \gamma_0^R L_R + c_1^{\perp} \bigg] \nonumber
\\
+ \left( \alspi \right)^2 \bigg[ \frac{\beta_0\Gcusp_0}{6} \Lp^3 + \left( \frac{\Gcusp_1}{2} - \frac{\beta_0\Gcusp_0}{2} L_R - \beta_0 \gamma_0^S \right) \Lp^2
\\
+ \left( \left( 2\beta_0 \gamma_0^R - \Gcusp_1 \right) L_R - 2\gamma_1^S + \beta_0 c_1^{\perp} \right) \Lp + 2\gamma_1^R L_R + c_2^{\perp} \bigg] \Bigg\} \, ,
\end{multline}
where $L_R = \Lp + L_{\nu}$ with $L_\nu = \ln(\nu^2/\mu^2)$ and the scale-independent terms are
\begin{align}
c_1^{\perp} &= -2 C_F \zeta_2 \, , \nonumber \\
c_2^{\perp} &= C_F C_A \left( -\frac{67}{3} \zeta_2 - \frac{154}{9} \zeta_3 + 10 \zeta_4 + \frac{2428}{81} \right) + C_F N_f \left( \frac{10}{3} \zeta_2 + \frac{28}{9} \zeta_3 -\frac{328}{81} \right).
\end{align}

\bibliographystyle{JHEP}
\bibliography{twoloop_TMD_quark}

\end{document}